\documentclass[lineno]{JFM-FLM_Au}
\usepackage{subfig}
\usepackage{amsmath}
\usepackage{graphicx}
\usepackage{float}
\usepackage{hyperref}
\usepackage{cleveref}

\lefttitle{Three-dimensional organisation and dynamics of stall cells}
\righttitle{Journal of Fluid Mechanics}

\title{ Stall cells over an airfoil. Part 1: Three-dimensional flow organisation and vorticity dynamics}

\author{Rishabh Mishra\aff{1}, Emmanuel Guilmineau\aff{1}, Ingrid Neunaber\aff{2}, and Caroline Braud \aff{1}}

\affiliation{\aff{1}LHEEA lab. - CNRS - Nantes Université, Centrale Nantes, 1 rue de la No\"e,
44100 Nantes, France
\aff{2} FLOW, Department of Engineering Mechanics, KTH Royal Institute of Technology, SE-100 44 Stockholm, Sweden}

\corresau{Emmanuel Guilmineau, \email{emmanuel.guilmineau@ec-nantes.fr}}

\begin{document}
\nolinenumbers
\maketitle

\begin{abstract}
This study investigates the three-dimensional organisation and evolution of stall cells in the separated flow region over an airfoil. Using a hybrid RANS/LES approach based on the DDES-SST turbulence model, we characterise the formation and development of these structures, which remain challenging to capture experimentally. Initial validation confirms accurate reproduction of global loads when comparing with both experimental data and RANS simulations. The complex three-dimensional flow organisation is analysed through investigating the vorticity, revealing that spanwise variation of the separation location leads to non-uniform load distribution along the airfoil span. The mid-span experiences premature separation due to flow bifurcation, while flow attraction at $\pm1$ chord length successfully sustains attached flow further along the chord. The separated flow generates a shear layer culminating in a separation vortex tube, which exhibits a Crow-type instability when interacting with the counter-rotating trailing edge vortex tube. This instability induces a wave-like bending of the vortex tubes and shear layer, generating significant vertical vorticity (y-vorticity) that drives spanwise flow. We identify a previously unreported phenomenon where the maxima of spanwise velocity structures exhibit rotation around fixed spanwise axes, with the rotation angle evolving linearly with downstream distance according to $\zeta = 14.5(x/c) - 0.8$. This study provides new insights into the mechanisms underlying stall cell formation and highlights the importance of three-dimensional effects in separated flows, which has implications for aerodynamic load prediction and control strategies. 
\end{abstract}

\newpage
\section{Introduction}

Separated flows over airfoils exhibit complex three-dimensional structures that significantly influence aerodynamic performance across various engineering applications. Among these structures, stall cells, organised spanwise patterns of alternating flow direction in separated regions, represent a fundamental mechanism affecting load distribution, stability, and aerodynamic performance. While trailing edge separation on two-dimensional airfoils is often treated as a largely two-dimensional phenomenon in simplified models, experimental evidence has consistently demonstrated its inherently three-dimensional nature, particularly through the formation of stall cells.

\subsection{Historical Development of Stall Cell Research}

\noindent Numerous experimental and numerical studies on airfoils have identified 3D, mushroom-shaped flow structures. The earliest reports date to the 1970s, notably \cite{moss1971two} and \cite{gregory1971progress}, in which wings spanned the full wind-tunnel width. They observed structures that were initially attributed to wind-tunnel wall effects. However, \cite{winkelman1980flowfield} performed oil-film measurements on rectangular wings with aspect ratios (AR) of 3, 5, 6, and 9 at chord-based Reynolds numbers between 245,000 and 385,000, that did not span the whole width of the wind tunnel, and their results showed that stall cells also occur with free tips sufficiently distant from the tunnel walls, thereby ruling out wind tunnel wall effects as the cause. They further observed that the number of stall cells increases with increasing AR. To the best of the author’s knowledge, \cite{Bertagnolio2005} were the first to report stall cells in a numerical study. They carried out Unsteady Reynolds Averaged Navier Stokes (URANS) simulation and Detached Eddy Simulation (DES) on a RISO-B1-18 airfoil at \(Re_c = 1.6\times10^6\) and detected stall-cell structures in their URANS computations. \noindent \cite{gross2011numerical} conducted several simulations of stalled airfoils--URANS, hybrid RANS/LES, and under-resolved DNS (implicit LES)—but the domain for the DNS (implicit LES) case was too small for a meaningful stall-cell analysis. Their 3D URANS results revealed a curved separation line and a marked reduction in lift coefficient relative to 2D simulations, indicating the presence of stall cells. Using RANS, \cite{kamenetskiy2014numerical} observed stall-cell-like structures and similarly obtained a lower lift coefficient than in 2D, reporting a stall-cell wavelength of one chord, limited by the computational span. Finally, \cite{manolesos2014experimental} used URANS with a zigzag tape to perturb the flow and showed that CFD can qualitatively reproduce stall-cell behaviour, with stall-cell widths ranging from 20\% to a full chord length.\\

\subsection{Vortex organisation}

\noindent The vortex structures associated with stall cells are outlined below (adapted from \cite{manolesos2014study}).

\begin{figure}[H]
  \centerline{\includegraphics[width=0.8\textwidth]{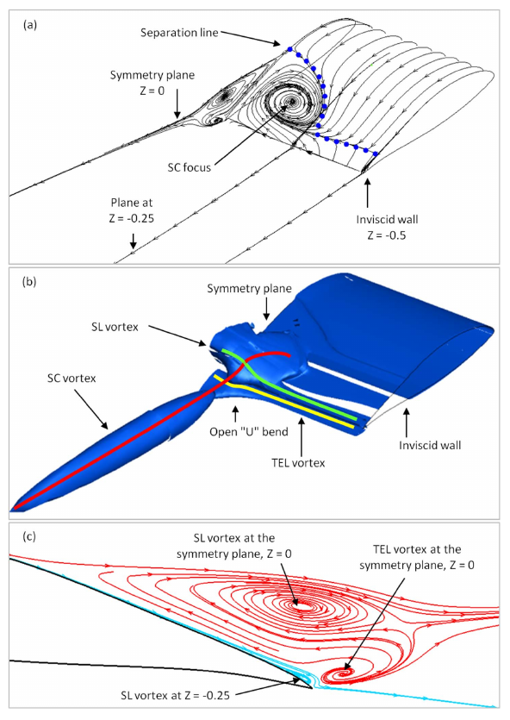}}
  \caption{visualisation of the CFD-derived stall cell (SC) structures. In these images, the separation line vortex is designated as SL vortex and the stall cell vortex as SC vortex. The flow direction is indicated by blue arrows. (a) Depicts the surface flow lines on the wing's suction surface, complemented by in-plane flow lines on planes $z= 0$ (left side—symmetry plane), $z = -0.25$, and $z = -0.5$ (right side—inviscid wall). The 3D separation line is marked by blue dots. (b) Features a Q = 1 iso-surface with highlighted vortex core lines indicating vorticity direction. The distinctive open “U” shape of the Trailing Edge Line Vortex (TELV) inboard of the SC vortices is also shown. (c) Provides a side view detailing in-plane flow lines on plane $z = 0$ (symmetry plane—red lines) and $z = -0.25$. Specifications include an aspect ratio of 2.0, Reynolds number $Re_c = 8.7 \times 10^5$, and angle of attack $\theta = 10^\circ$ (adapted from \cite{manolesos2014study} with permission from AIP publising under licence number: 6202370215761).}
\label{fig:stall cell vortex organisation}
\end{figure}

\begin{enumerate}
    \item \textbf{Counter-rotating stall-cell vortices:} These vortices originate nearly normal to the wing surface and convect downstream; their cores lie within the wing’s separation line (see figure \ref{fig:stall cell vortex organisation} (a)).
    \item \textbf{Separation-line vortex (SL vortex):} Emanating from the stall cell centre, this vortex runs along—and slightly above—the wing’s trailing edge, extending laterally beyond the stall cell while remaining approximately parallel to the trailing edge (see figure \ref{fig:stall cell vortex organisation} (b)).
    \item \textbf{Trailing-edge line vortex (TEL vortex):} Rotating opposite to the SLV, the TELV forms parallel to the trailing edge at the stall cell’s outer margins and grows inboard toward the cores of the stall-cell vortices (see figure \ref{fig:stall cell vortex organisation} (c)).
\end{enumerate}

\subsection{Impact of aspect ratio}

\noindent \cite{schewe2001reynolds} reported two cells for an aspect ratio (AR) of 4 and four cells for an AR of 6 on a thick airfoil, indicating that the number of stall cells scales with AR, with higher-AR airfoils exhibiting more cells. \cite{zarutskaya2005vortical} presented half-wing RANS simulations using the Spalart--Allmaras turbulence model (see \cite{spalart2007effective}) on a NACA0012 at $Re_c = 3.9\times10^5$ and $Re_c = 4.6\times10^6$, and on NASA SC2 airfoils with AR = 8 at $Re_c = 4.6\times10^5$. For all three cases, a symmetry boundary condition was imposed on one side and a free-tip boundary condition on the other, and two stall cells were observed. \noindent \cite{elimelech2012three} used constant-temperature anemometry (CTA) in the vicinity of the airfoil surface to examine two airfoils at Reynolds numbers between 10,000 and 20,000, identifying two and four cellular patterns for ARs of 2.5 and 5, respectively. \noindent \cite{manni2016numerical} simulated a high-AR wing using URANS and DDES in \textsc{Fluent}, observing stall cells with sizes between $1.4c$ and $1.7c$, and noting that lift coefficients were lower when stall cells were present than in corresponding 2D simulations. 

\subsection{Low-frequency oscillations}
\noindent \cite{yon1998study} investigated a NACA 0015 airfoil at a Reynolds number of $6.2 \times 10^5$, assuming a turbulent boundary layer on the suction side. They reported a stable pattern featuring two cells for aspect ratios between 3.5 and 4.5, indicating a preferred aspect ratio. High-amplitude, low-frequency pressure fluctuations measured with pressure sensors were observed within the stall cells, with a Strouhal number of 0.04, consistent with \cite{zaman1989natural}.\\
\cite{liu2018numerical} performed 3D unsteady RANS simulations on a NACA 0012 airfoil, varying Reynolds number, angle of attack, chord-to-span ratio, and spanwise mesh resolution. The results show a clear correlation between stall-cell oscillation and lift fluctuation, particularly in the post-stall region between $17^\circ$ and $19.5^\circ$ at a chord-based Reynolds number of $10^6$. Proper orthogonal decomposition (POD) revealed that the first few modes correspond to stall-cell profiles and correlate with lift fluctuations.\\
\cite{bouchard2022numerical} examined the flow around a stalled airfoil using zonal detached-eddy simulation with transition effects. The study revealed mixed trailing-edge/leading-edge stall characteristics at a chord-based Reynolds number of $10^6$ and a Mach number of 0.16. Two computations with different initial conditions showed that starting from attached flow led to stable attachment, whereas starting from separated flow produced low-frequency oscillatory behaviour and reattachment after 90 chord-passing durations. During the oscillatory phase, stall cells emerged and vanished, with low-frequency flapping at the leading edge driven by high-frequency fluctuations originating from the trailing edge. The low-frequency oscillation thus appears to coincide with the formation of stall cells. However, this contradicts the findings of \cite{broeren2001spanwise,broeren1998low}, who demonstrated that, depending on the stall type, either low-frequency oscillation or stall cells are observed, but not both simultaneously. 
\cite{hanna2026trackingstallcelldynamics} recently demonstrated that low-frequency oscillations are localised around stall cell borders, a phenomenon that global load measurements fail to capture. These large-amplitude fluctuations, previously observed by \cite{Neunaber2022} and \cite{BraudPhysreview2024}, exhibit a high degree of spatial anti-correlation in the spanwise direction, resulting in the wall pressure bi-stability that drives significant load fluctuations. This behavior was subsequently linked by \cite{hanna2026trackingstallcelldynamics} to the unsteady spanwise displacement of the stall cell. At Reynolds numbers exceeding $10^6$, the dynamics are dominated by a coherent spanwise motion with a characteristic velocity of approximately $0.1U$. This motion is characterised by a large-scale, low-frequency sweep with a Strouhal number $St \sim 0.001$, which is superimposed with smaller-scale, higher-frequency oscillations.

\subsubsection{Angle of attack, Reynolds number, surface roughness, and camber}

\noindent \cite{sarlak2018experimental} examined an S826 airfoil and showed that stall-cell formation depends on Reynolds number: no cells appeared at low Reynolds numbers, whereas adding surface roughness at low Reynolds numbers produced more distinct vortex pairs.\\
\noindent \cite{yon1998study} identified stall cells within a narrow angle-of-attack band of $2^\circ-3^\circ$ centred around $17^\circ$ for a NACA 0015 airfoil at $Re_c = 6.2 \times 10^5$. Stall cells were also observed by \cite{broeren2001spanwise} on ‘Ultra-Sport’ and NACA 2414 airfoils over AoAs near the angle of maximum lift at \(Re_c = 3.0 \times 10^5\).\\
\noindent \cite{dell2016measurement} and \cite{dell2018parametric} reported that stall cells on a NACA 0015 airfoil occur only for specific combinations of Reynolds number and AoA, identifying eight distinct surface oil-flow patterns and inducing stall cells using zigzag tape. An additional investigation by \cite{de2022effects} on the NACA 0012 airfoil revealed two Reynolds-number regimes: below $Re_c = 1.5 \times 10^5$, the transition from separated flow to stall cells is primarily governed by changes in $Re_c$; above $Re_c = 1.5 \times 10^5$, the transition is mainly controlled by changes in AoA. The study further indicated that, for $Re_c > 1.5 \times 10^5$, reducing camber delays stall-cell onset to higher AoAs.\\
\noindent \cite{manolesos2014geometrical}, \cite{manolesos2014experimental}, and \cite{manolesos2015experimental} investigated a wind-turbine airfoil exhibiting trailing-edge stall. Using zigzag tapes to impose local or full-span disturbances, they observed multiple stall-cell patterns, including 1, 1.5, and 2 cells for $AR = 2$.\\
\noindent \cite{hanna2026trackingstallcelldynamics} investigated stall cells over a thick airfoil (20\%) at high Reynolds numbers (from $0.5 \times 10^6$ to $3.4 \times 10^6$) and angles of attack within the stall cell regime, spanning from 12° to 16°. They found that stall cell width increases almost linearly with the angle of attack, while the stall cell is split into two parts at Reynolds numbers below $10^6$. In that study, the stall cell regime is marked by large local wall pressure fluctuations whose amplitude increases significantly with the Reynolds number, indicating that this local phenomenon is characteristic of high Reynolds numbers. 

\subsection{Impact of numerical parameters}

\noindent The spanwise extent of the computational domain must be large enough to permit the formation of at least one stable stall cell (see \cite{manolesos2021investigation}). Numerical and experimental evidence indicates that the domain width should be no less than one chord length (see, e.g., \cite{weihs1983cellular,manni2016numerical}). Moreover, \cite{spalart1997sensitization} showed that the spanwise mesh spacing is constrained by turbulence-resolution requirements, necessitating a large number of cells in the spanwise direction. In addition, the number of time steps must be sufficiently high to capture low-frequency oscillations (LFOs) effectively (see \cite{deck2014zonal}).

\noindent \cite{liu2018numerical} performed 3D unsteady RANS simulations on a NACA 0012 aerofoil, varying Reynolds number, angle of attack, chord-to-span ratio, and spanwise mesh resolution. 
They found that unsteadiness was pronounced with a medium spanwise resolution of mesh (10\% of the chord) and more moderate with finer resolutions (5\% and 2.5\% of the chord).

\noindent From the studies cited in this section, it follows that capturing phenomena such as LFOs requires finer mesh resolution to mitigate unsteadiness, together with a sufficiently large number of time steps. Conversely, accurately reproducing the correct number of stall cells depends primarily on selecting an appropriate domain span.

\subsection{Modelling of stall cells}

\begin{figure}[H]
  \centerline{\includegraphics[width=\textwidth]{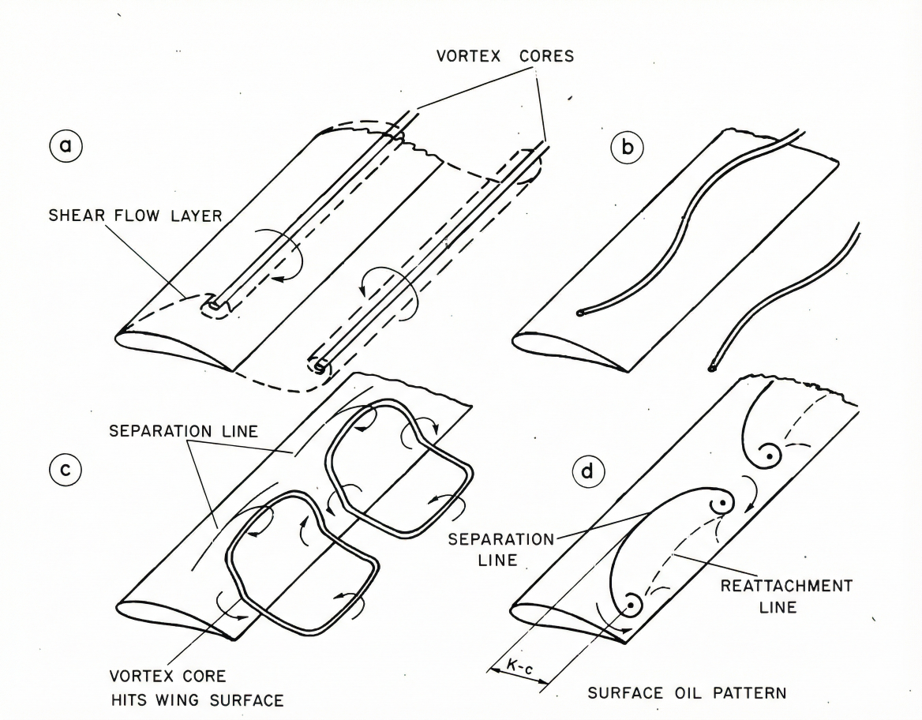}}
  \caption{Diagram illustrating vortex instability and the formation of cellular patterns in the separated flow over rectangular wings (taken from \cite{weihs1983cellular}).}
\label{fig:crow_instability_diagram}
\end{figure}

\noindent The earliest modelling effort for stall cells is due to \cite{weihs1983cellular}, who proposed their occurrence via a Crow-type instability (see \cite{crow1970stability}). The model assumes a spanwise vortex line on the suction side of the airfoil (see figure \ref{fig:crow_instability_diagram}). Based on this, the stall-cell wavelength was given as $\Lambda_c = 2h/8.6$, where $h$ is the distance from the vortex core to the suction surface measured along the line normal to the chord. This model matched well the dataset used in their study and showed good agreement with the experiments of \cite{winkelman1980flowfield}. However, substantial discrepancies were reported relative to URANS results from \cite{gross2011numerical}.\\
\noindent \cite{rodriguez2011birth} interpreted stall cells as arising from a spanwise-unstable global mode of separation in the laminar regime, which renders a nominally 2D airfoil flow three-dimensional. This mechanism applies only to cases with laminar separation.\\
\noindent \cite{spalart2014prediction} and \cite{gross2015criterion} introduced inviscid, theory-based criteria, both indicating that a negative slope of the $C_l$ curve is a necessary condition for stall-cell onset. Further study is required, as this conclusion does not align with certain observations—see, for example, \cite{broeren2001spanwise}.

\begin{table}
  \centering
  \small
  \def~{\hphantom{0}}
  \begin{tabular}{@{}lcccccc@{}}
    \textbf{Reference }& \textbf{E/N} & \textbf{Airfoil} & \textbf{$Re_c$} & \textbf{AR} & $\Lambda/c$ & $N_{\mathrm{SC}}$ \\
    \cite{moss1971two} & E & NACA 0012 & $0.86$--$1.68 \times 10^6$ & 3.75& 2.5 & ${\sim}1.5$ \\
    \cite{gregory1971progress} & E & NACA 0012 & $1.7$--$0.85 \times 10^6$ & 1.4--2.8 & 2.5 & ${\sim}1$ \\
     &  & NPL96192 & $0.76 \times 10^6$ & 6 & -- & -- \\
     &  & NPL96192 & $2.88 \times 10^6$ & 3.59 & ${\sim}1.2$ & ${\sim}3$ \\
     &  & NPL96192 & $3.14 \times 10^6$ & 3.29 & ${\sim}1.6$ & ${\sim}2$ \\[3pt]
    \cite{winkelman1980flowfield} & E & Clark Y & $2.5$--$3.9 \times 10^5$ & 3, 6, 9 & 3 & 1, 2, 3 \\[3pt]
    \cite{yon1998study} & E & NACA 0015 & $6.2 \times 10^5$ & 2--6 & 1.5--2.3 & 2 \\
    \cite{schewe2001reynolds} & E & FX-77-W270 & $0.32$--$10.0 \times 10^6$ & 4 & 2 & 2 \\
     &  &  &  & 6 & -- & 4 \\
    \cite{broeren2001spanwise} & E & Ultra-Sport & $3.0 \times 10^5$ & 2.8 & 2 & ${\sim}1$ \\
     &  & NACA 2412 &  &  &  &  \\
     &  & NACA 64A010 &  &  &  &  \\
     &  & LRN-1007 &  &  &  &  \\
     &  & E374 &  &  &  &  \\[3pt]
    \cite{Bertagnolio2005} & N & RIS{\O}-B1-18 & $1.6 \times 10^6$ & 2.56 & ${\sim}1.25$ & ${\sim}2$ \\
    \cite{zarutskaya2005vortical} & N & NACA 0012 & $3.9 \times 10^5$ & 8 & ${\sim}1.5$ & ${\sim}5$ \\
     &  & NASA SC2 & $4.6 \times 10^6$ &  &  &  \\
    \cite{gross2011numerical} & N & NACA $64_3$-618 & $0.322$--$2.63 \times 10^6$ & 0.4, 0.6 & -- & -- \\
    \cite{elimelech2012three} & E & NACA 0009, Eppler-61 & $1$--$2 \times 10^4$ & 2.5--5 & 1.5 & 2 \\
     &  &  &  & 5 & -- & 4 \\[3pt]
    \cite{kamenetskiy2014numerical} & N & NACA 0012 & $5$--$18 \times 10^6$ & 1 & $\sim 1$ & ${\sim}1$ \\
    \cite{manolesos2014geometrical} & E & NTUAt18 & $0.5$--$1.5 \times 10^6$ & 1.5, 2.0 & 1 & 1.5, 2 \\
    \cite{manolesos2014experimental} & N, E & NTUAt18 & $1 \times 10^6$ & 2 & ${\sim}1.25$ & 1, 1.5, 2 \\
    \cite{manolesos2015experimental} & E  & NTUAt18 & $0.87 \times 10 ^6$  & -- & -- &  \\
    \cite{manni2016numerical} & N & High-AR wing & -- & High & 1.4--1.7 & -- \\
    This work &N& Wind turbine & $0.2 \times 10^6$ &4,6 &2, 1.5 &2,3\\
  \end{tabular}
    \caption{Summary of experimental and numerical studies on stall cells. Here, $Re_c$ denotes the chord-based Reynolds number, AR the aspect ratio, $\Lambda$ the stall-cell wavelength or width, and $N_{\mathrm{SC}}$ the number of stall cells observed. E = Experimental, N = Numerical. A dash (--) indicates that the parameter was not reported or is not applicable.}
  \label{tab:stall_cells_summary}
\end{table}

\subsection{Knowledge gaps and present contribution}

Despite significant progress in understanding stall cells, several important knowledge gaps remain. Most previous studies have focused on low turbulence intensity inflows, leaving open questions about stall cell behavior under the more turbulent conditions typical of real-world applications such as wind turbine blades. Additionally, the quantitative characterisation of the streamwise evolution of stall cell structures remains limited, with few studies tracking their development downstream of the airfoil.

The vorticity dynamics underlying stall cell formation, while increasingly studied, have not been fully characterised in terms of the interaction between different vorticity components and their role in establishing and maintaining the three-dimensional organisation of the flow. Furthermore, the relationship between global stability characteristics predicted by linear theory and the nonlinear vorticity dynamics observed in fully developed stall cells remains incompletely understood.

Most numerical studies have employed either simplified RANS models, which struggle to capture the full three-dimensional dynamics, or computationally intensive LES approaches, which are difficult to apply to practical problems. The development and validation of intermediate fidelity approaches, such as  hybrid RANS/LES approach , partuclarly delayed detached eddy simulation (DDES), represents an important area for ongoing research.

The present work aims to address these knowledge gaps through high-fidelity numerical simulations using a hybrid RANS/LES approach. Specifically, we seek to:

\begin{enumerate}
\item characterise the three-dimensional organisation of stall cells and their impact on local load distribution under moderate turbulence conditions ($TI = 3\%$)
\item Elucidate the vorticity dynamics underlying stall cell formation, particularly the interaction between spanwise, streamwise, and vertical vorticity components
\item Quantify the streamwise evolution of stall cell structures, identifying and explaining previously unreported patterns in their downstream development
\end{enumerate}

This manuscript is organised as follows: Section \ref{sec:s2} describes the airfoil used for the study, the numerical methodology, including the DDES-SST turbulence model and computational setup. Section \ref{sec:s3} presents a validation of the approach through global load assessment. Section \ref{sec:s4} analyses the local load distribution and its relationship to stall cell structure. Section \ref{sec:s5} examines the wall friction and mean kinetic energy distribution to characterise separation patterns. Section \ref{sec:s6} identifies stall cell structures through flow visualisation. Section \ref{sec:s7} presents a detailed analysis of the mean flow field and vorticity dynamics.  Section \ref{sec:s8} investigates the effect of the increasing AoA on the stall cells.  Finally, Section \ref{sec:s9} concludes and summarises the key findings and their implications.

The physical mechanisms identified in the present work provides the 
foundation for a vortex-based analytical model developed in a companion 
paper \cite{PartII}. That work extends classical vortex stability theory 
through a weakly nonlinear analysis to derive the saturation amplitude, 
vortex sheet dynamics, and the induced spanwise velocity field, with 
predictions validated against the simulation results presented herein.

\section{Numerical methodology} \label{sec:s2}

\subsection{Airfoil description}

\begin{figure}
\centerline{\includegraphics[width=0.8\textwidth]{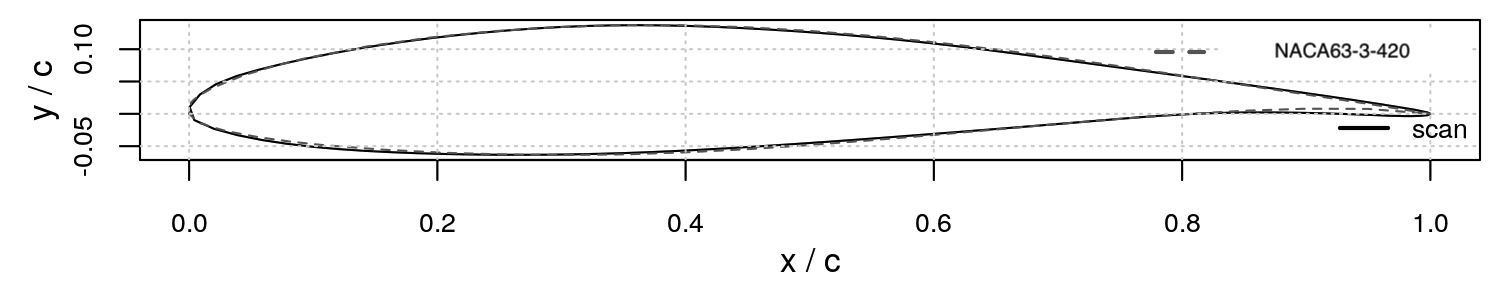}}
  \caption{Blade section at 82\% of the radius of a full-scale wind turbine in comparison with a NACA63-3-420 profile with a modified camber of 4\% instead of 2\% (from \cite{Neunaber2022}; published under CC BY-NC-ND 4.0.)}
\label{fig:Profile1}
\end{figure}

The airfoil shape used as reference in this study was derived from scans of a 2MW wind turbine blade, at 82$\%$ of its length (\cite{Neunaber2022}). It has been scaled-down to 1/10th of the original chord length, so that the chord length is $c=0.125$~m and the chord-based Reynolds number is $Re_c=2.0 \times 10^5$. The airfoil section closely resembles a NACA63-3-420 profile with a modified camber of 4\% instead of 2\% (see figure \ref{fig:Profile1}).  

\subsection{Computational approach}

To capture the three-dimensional organisation of stall cells, which are difficult to fully characterise experimentally, we employed a hybrid RANS/LES approach based on the Delayed Detached Eddy Simulation (DDES) with the Shear Stress Transport (SST) turbulence model \citep{gritskevich2012development}. This DDES-SST model strikes an effective balance between computational efficiency and resolution of high-energy flow structures, making it well-suited for studying the complex three-dimensional separated flows associated with stall cells.

The DDES approach functions as a RANS model near walls, where the boundary layer is thin and grid refinement requirements for LES would be prohibitive, while transitioning to an LES-like behavior in separated regions and away from walls. This hybrid nature allows for accurate resolution of the large-scale vortical structures that characterise stall cells while maintaining reasonable computational costs. All simulations were run using the ISIS-CFD solver developed by CNRS and Centrale Nantes. A brief introduction to this solver is given in the next section.

\subsection{ISIS-CFD}

The ISIS-CFD solver, developed jointly by CNRS and Centrale Nantes, is part of the FINE\texttrademark/Marine software suite, which is distributed worldwide by Cadence Design Systems. This solver employs an incompressible unsteady Reynolds-averaged Navier–Stokes (URANS) formulation. The transport equations are spatially discretised using the finite volume method (see \cite{moukalled2016finite}). The unstructured, face-based discretisation supports cells with an arbitrary number of faces of arbitrary shape. Temporal discretisation is achieved through a second-order backward difference scheme. All flow variables are stored at the geometric centres of the cells, while surface and volume integrals are evaluated with second-order accuracy. Being face-based, the method reconstructs numerical fluxes on mesh faces via linear extrapolation of the integrand from adjacent cell centres. A centred scheme is used for diffusive terms, whereas convective fluxes are computed using a blended approach comprising 80\% central and 20\% upwind differencing. For turbulent flows, supplementary transport equations corresponding to the turbulence model variables are also solved.

\subsection{Computational domain and mesh}

\begin{figure}
\centering
\includegraphics[width=1.0\textwidth]{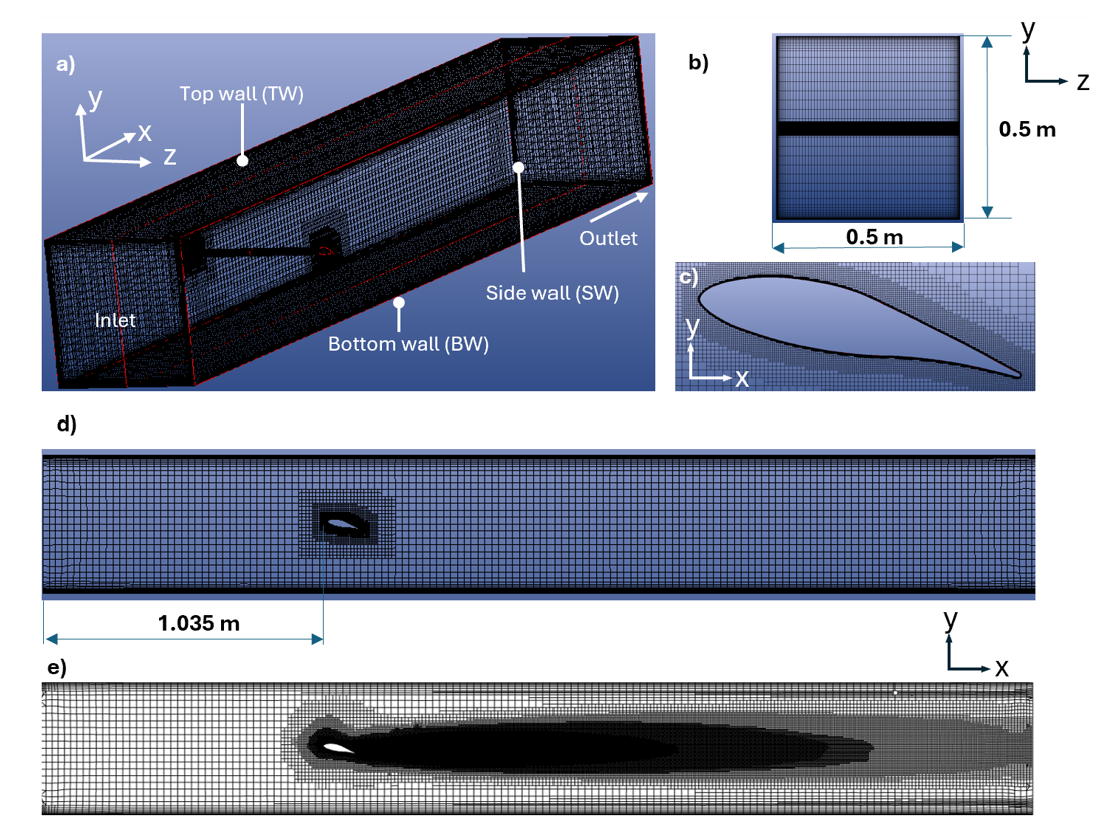}
\caption{Views of the mesh: (a) 3D mesh for the simulation of the flow over the airfoil (13 million cells); (b) side view of the mesh; (c) mesh around the airfoil; (d) front view of the mesh; (e) mid span view of the mesh at the end of the simulation after adaptive grid refinement (130 million cells). The figure is adopted from \cite{mishra2024developing}, published under CC BY 4.0.}
\label{fig:mesh_views_full}
\end{figure}

Figure \ref{fig:mesh_views_full} shows the three-dimensional view, front view, and side view of the numerical domain, respectively, for the simulation of flow over an airfoil at 14° angle of attack. The full cross-section was modeled here to enable the simulation of complete three-dimensional flow, particularly the spanwise organisation of stall cells. The baseline mesh consisted of 13 million cells, with the airfoil having an aspect ratio of 4, corresponding to the experimental configuration.

Figure \ref{fig:mesh_views_full} also presents a close-up view of the mesh around the airfoil, showing the refinement near surfaces to capture boundary layer development accurately. To enhance computational efficiency while maintaining resolution in critical regions, an adaptive grid refinement methodology was employed \citep{wackers2012adaptive}. This approach dynamically increased mesh resolution in regions of high flow gradients, particularly in the separated flow region downstream of the airfoil. By the end of the simulation, the adaptive refinement had increased the total cell count to approximately 130 million cells.

\subsection{Boundary conditions and numerical setup}

 We have applied a Dirichlet boundary condition at the inlet, and the values are given in table \ref{tab:BC at inlet}. For pressure, the Neumann boundary condition was applied at the inlet, $\frac{dp}{d\mathbf{n}}=0$, where $\mathbf{n}$ is the normal vector to the inlet. These same values have been used as the initial conditions as well. The Dirichlet boundary condition was applied at the outlet for pressure $p =p_o$, where $p_o = 0$ by default. For TKE and turbulence frequency, the Neumann boundary condition was applied as $\frac{dk}{d\mathbf{n}} = 0$ and $\frac{d\omega}{d\mathbf{n}} = 0$, respectively. A no-slip boundary condition was applied on the airfoil, and wall functions,
\begin{equation}
    \frac{\partial U}{\partial y} = \frac{\tau_w}{\kappa \rho c^{1/4}_\mu \sqrt{k_w} y_w},
    \label{eq: wall law}
\end{equation}
 were imposed on the top wall (TW), bottom wall (BW), and side wall (SW) to avoid explicitly simulating the boundary layer. Here, $U$ is the velocity, $k_w$ is the TKE at the cell centre of the first cell from the wall, $y_w$ is the perpendicular distance of the cell centre of the first cell from the wall, $\tau_w$ is the wall shear stress, $\kappa = 0.41$, and $c_\mu = 0.09$.

 The time step for the simulation is $5 \times 10^{-5}$ s. For unsteady simulations data was recorded at every 20th time step.

Table \ref{tab:y+ values} presents the $y+$ values applied and achieved in the simulation. The airfoil surface was resolved with a fine mesh yielding an average $y+$ of 0.039, ensuring accurate resolution of the boundary layer development and separation process. The tunnel walls were treated with wall functions and maintained appropriate $y+$ values to capture their influence on the flow field without excessive computational cost.

\begin{table}
\centering
\def~{\hphantom{0}}
\begin{tabular}{c c }
\textbf{Variable} & \textbf{Value at inlet}\\
$U$      & $25$~ms$^{-1}$ \\
$k$      & $1.859$~m$^{2}$s$^{-2}$\\
$\omega$ & $657.4$~s\textsuperscript{-1}  \\
$L_S$    & $2.54$~mm   \\
\end{tabular}
\caption{Boundary conditions at the simulation domain inlet. Note that for the inlet length scale $L_S$, the Taylor micro-scale is used.}\label{tab:BC at inlet}
\end{table}

\begin{table}
\centering
\def~{\hphantom{0}}
\begin{tabular}{l c c c }
\textbf{Boundary} & \textbf{Applied $y^+$} & \textbf{Average $y^+$} & \textbf{$y^+$ Range} \\ 
Airfoil           & 0.15                   & 0.039                   & 0.02 - 0.16          \\
Top Wall (TW)     & 50                     & 13                     & 0.6 - 20             \\ 
Bottom Wall (BW)  & 50                     & 13                     & 0.6 - 20             \\
Side Walls (SW)    & 1                      & 1                      & 0.2 - 12            \\ 
\end{tabular}
\caption{$y^+$ values for the simulation.}\label{tab:y+ values}
\end{table}

Due to the high computational expense of DDES simulations, they were performed for selected angles of attack (10°, 14°, 16°, and 20°), with particular focus on the 14° case which lies in the transition-to-stall region where complex three-dimensional flows induced by trailing edge separation can be observed. The simulations employed a blended numerical scheme with an upwinding level of 0.05 to balance accuracy and stability.

In all subsequent analyses, results are presented in the airfoil coordinate system, normalized by the chord length $c$. The origin is located at the leading edge of the airfoil at 0° angle of attack. Before, going further a brief description of DDES-SST is give in the next section.

\subsection{DDES-SST}

\noindent The DDES-SST model belongs to the class of hybrid RANS/LES approaches. In this framework, the RANS formulation is employed in the near-wall regions where a fully resolved LES would be computationally too demanding. Away from the wall, the model switches to LES to better capture the large-scale turbulent eddies. The transition between the RANS and LES zones is governed dynamically by a blending function that depends on both the local flow features and the grid resolution, ensuring a smooth and consistent shift from RANS to LES (\cite{gritskevich2012development}).\\

\noindent The governing transport equations of the DDES-SST model are expressed as (adapted from \cite{gritskevich2012development}):

\begin{equation}
\begin{aligned}
\frac{\partial \rho k}{\partial t} + \nabla \cdot (\rho \mathbf{u} k) &= \nabla \cdot \left[(\mu + \sigma_k \mu_t) \nabla k\right] + P_k - \rho \sqrt{k^3}/l_{\text{DDES}}, \\
\frac{\partial \rho \omega}{\partial t} + \nabla \cdot (\rho \mathbf{u} \omega) &= \nabla \cdot \left[(\mu + \sigma_\omega \mu_t) \nabla \omega\right] + 2(1-F_1)\rho\sigma_{\omega}^2 \frac{\nabla k \cdot \nabla \omega}{\omega} \\
&\quad + \alpha \frac{\rho \mu_t P_k}{k} - \beta \rho \omega^2.
\label{eq:DDES SST equation}
\end{aligned}
\end{equation}

\noindent In equation (\ref{eq:DDES SST equation}), SST blending functions are represented by $F_1$ and $F_2$. They are defined as
\begin{equation}
F_1 = \tanh \left(\arg_1^4\right),
\end{equation}
where
\begin{equation}
\arg_1 = \min \left(\max \left(\frac{\sqrt{k}}{C_\mu \omega d_w}, \frac{500 \nu}{d_w^2 \omega}\right), \frac{4 \rho \sigma_{\omega2} k}{CD_{k\omega} d_w^2}\right),
\end{equation}
and
\begin{equation}
CD_{k\omega} = \max \left(2 \rho \sigma_{\omega2}\frac{\nabla k \cdot \nabla \omega}{\omega}, 10^{-10}\right).
\end{equation}
\begin{equation}
F_2 = \tanh \left(\arg_2^2\right),
\end{equation}
with
\begin{equation}
\arg_2 = \max \left(\frac{2 \sqrt{k}}{C_\mu \omega d_w}, \frac{500 \nu}{d_w^2 \omega}\right).
\end{equation}

\noindent Following gives the production term of the equation (\ref{eq:DDES SST equation})
\begin{equation}
P_k = \min \left(\mu_t S^2, 10 \cdot C_\mu \rho k \omega\right).
\end{equation}

\noindent The DDES length scale, $l_{\text{DDES}}$, is defined as
\begin{equation}
l_{\text{DDES}} = l_{\text{RANS}} - f_d \max(0, l_{\text{RANS}} - l_{\text{LES}}),
\label{eq:DDES length blending}
\end{equation}
where $l_{\text{LES}} = C_{\text{DES}} h_{\text{max}}$, $l_{\text{RANS}} = \frac{\sqrt{k}}{C_\mu \omega}$, and $C_{\text{DES}} = C_{\text{DES1}} F_1 + C_{\text{DES2}} (1 - F_1)$. Here, $h_{\text{max}}$ denotes the maximum edge length of a computational cell.\\

\noindent The empirical blending function $f_d$ in equation (\ref{eq:DDES length blending}) is expressed as
\begin{equation}
f_d = 1 - \tanh \left[\frac{C_{d1} r_d}{C_{d2}}\right],
\end{equation}
where
\begin{equation}
r_d = \frac{\nu_t + \nu}{\kappa^2 d_w^2 \sqrt{0.5 (S^2 + \Omega^2)}}.
\end{equation}

\noindent Here, $S$ and $\Omega$ denote the magnitudes of the mean strain rate tensor and the mean vorticity tensor, respectively.\\

\noindent The model constants are summarised in Table~\ref{tab:ddes_sst_constants}. All other constants retain the same values as those defined for the $k$--$\omega$ SST model by \cite{menter2003ten}.

\begin{table}
\centering
\def~{\hphantom{0}}
\begin{tabular}{lccccccc}
\textbf{Constant}  &  $C_\mu$ & $\kappa$ & $a_1$ & $C_{\text{DES1}}$ & $C_{\text{DES2}}$ & $C_{d1}$ &   $C_{d2}$      \\ 
\textbf{Value} & 0.09  & 0.41 & 0.31 & 0.78  & 0.61  & 20 & 3              \\ 
\end{tabular}
\caption{Constants used for the DDES-SST model.}
\label{tab:ddes_sst_constants}
\end{table}

\section{Global load assessment} \label{sec:s3}

Before analysing the three-dimensional flow structures, we first validated our numerical approach by comparing the force coefficients obtained from the DDES-SST simulations with those obtained from experiments and from RANS simulations using the $k-\omega$ SST Menter 2003 model (see \cite{menter2003ten}) . For more information on experiments and RANS simulations refer to  \cite{mishra2024developing}.

Figure \ref{fig:Cl, Cd, menter vs DDES} compares the lift coefficient ($C_l$) and drag coefficient ($C_d$) values obtained from both simulation approaches. The angles of attack for DDES-SST simulations (10°, 14°, 16°, and 20°) were deliberately chosen to capture the transition-to-stall region where complex three-dimensional flows induced by trailing edge separation are expected to develop. This region begins at the end of the linear portion of the $C_l$ curve and extends to the stall angle of 20° at the Reynolds number and turbulence intensity conditions of this study.

\begin{figure}
\centering
\subfloat[]{\includegraphics[width=0.5\textwidth]{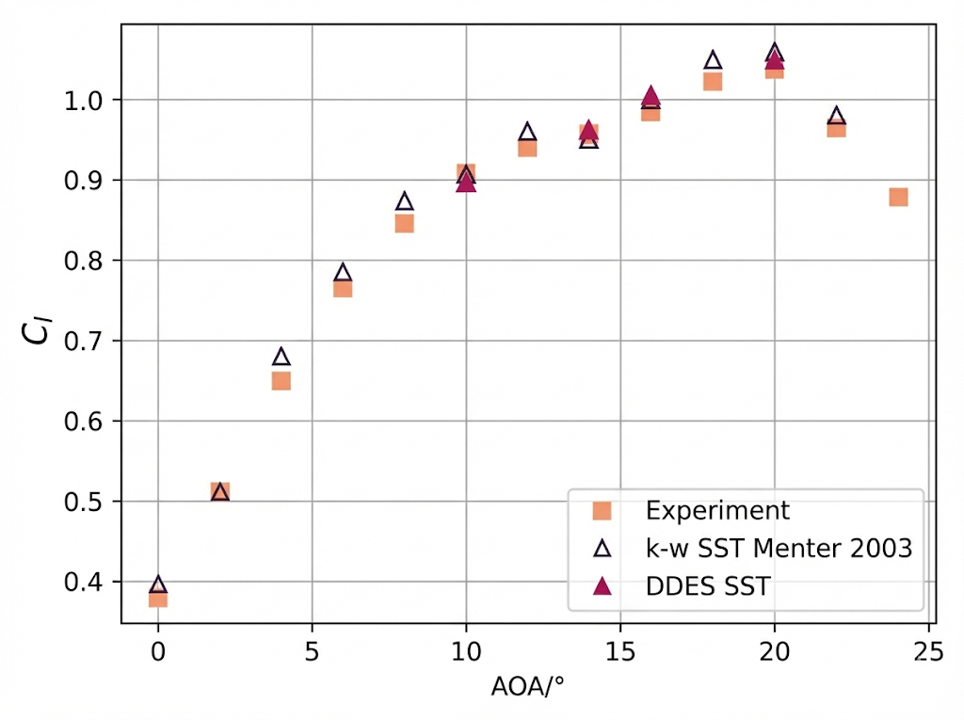}
\label{fig:Cl menter vs DDES}}
\subfloat[]{\includegraphics[width=0.5\textwidth]{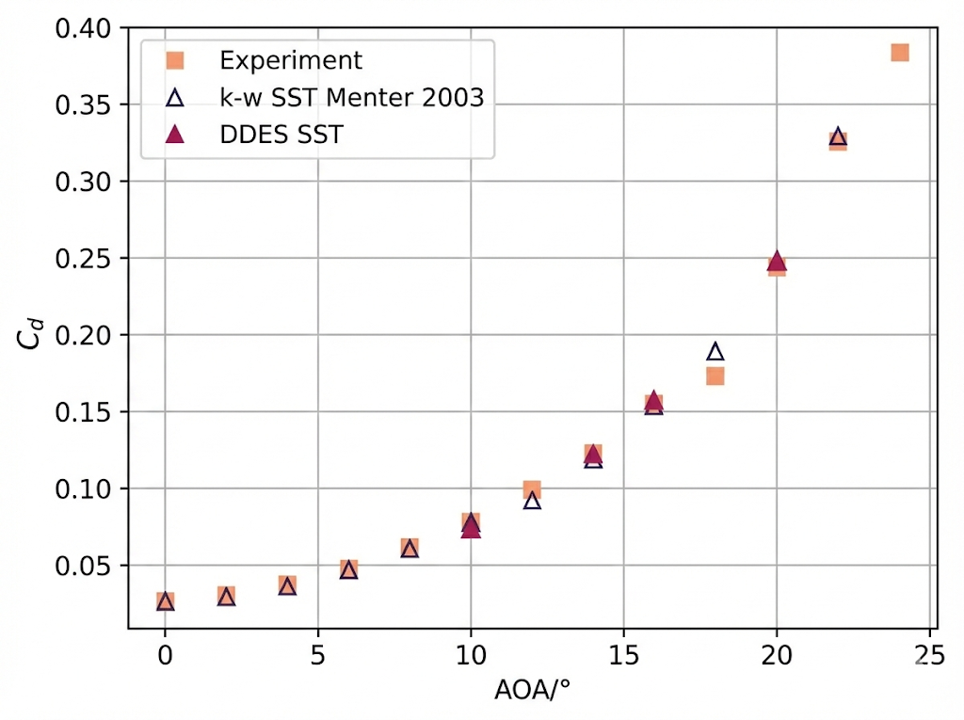}
\label{fig:Cd menter vs DDES}}
\caption{Comparison of (a) $C_l$ and (b) $C_d$ curves obtained using the DDES-SST turbulence model with experimental data and simulation result from the $k-\omega$ SST Menter 2003 model. The experimental and $k-\omega$ SST Menter 2003 model simulation data are taken from \cite{mishra2024developing}.}
\label{fig:Cl, Cd, menter vs DDES}
\end{figure}

The comparison shows excellent agreement between the force coefficients obtained using both models and the experiment. This alignment builds confidence in the DDES-SST approach and suggests that if the primary objective is to obtain global force coefficients, the computationally less expensive $k-\omega$ SST Menter 2003 model would be sufficient. However, for detailed investigation of the three-dimensional flow structures and vorticity dynamics that characterise stall cells, the higher-fidelity DDES-SST approach is necessary.

Having validated the global loads, the remainder of this manuscript focuses on detailed analysis of the three-dimensional flow organisation and stall cell structure, primarily using results from the 14° angle of attack case. This angle was specifically selected because stall cells are anticipated in thick trailing edge separation airfoils within the range of angles of attack where the $C_l$ curve flattens \citep{broeren2001spanwise}. For the same airfoil shape, this angle of attack was also found experimentally to exhibit bi-stability at a higher Reynolds number \citep{BraudPhysreview2024}.

\section{Local load distribution} \label{sec:s4}

\begin{figure}
\centering
\includegraphics[width=\textwidth]{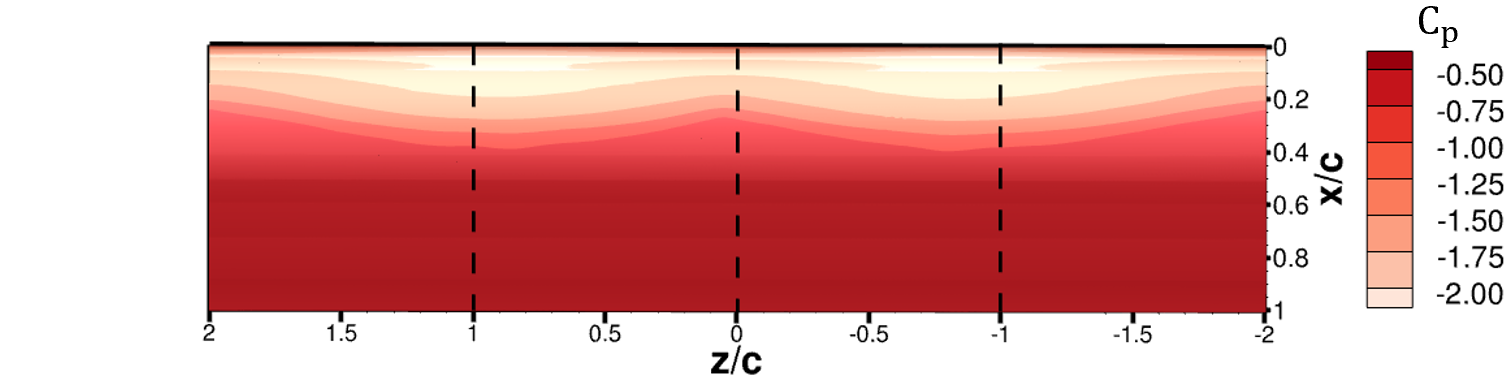}
\caption{$C_p$ distribution on the suction side of the airfoil}
\label{fig:Cp distribution on the suction side of the airfoil}
\end{figure}

The three-dimensional organisation of the flow field leads to significant spanwise variations in the load distribution. Figure \ref{fig:Cp distribution on the suction side of the airfoil} presents a contour plot of the time-averaged pressure coefficient ($C_p$) on the suction side of the airfoil, revealing two key features: (1) a non-uniform distribution in the spanwise direction, and (2) symmetry in the spanwise distribution around the $z/c = 0$ plane.

\begin{figure}
\centering
\includegraphics[width=0.7\textwidth]{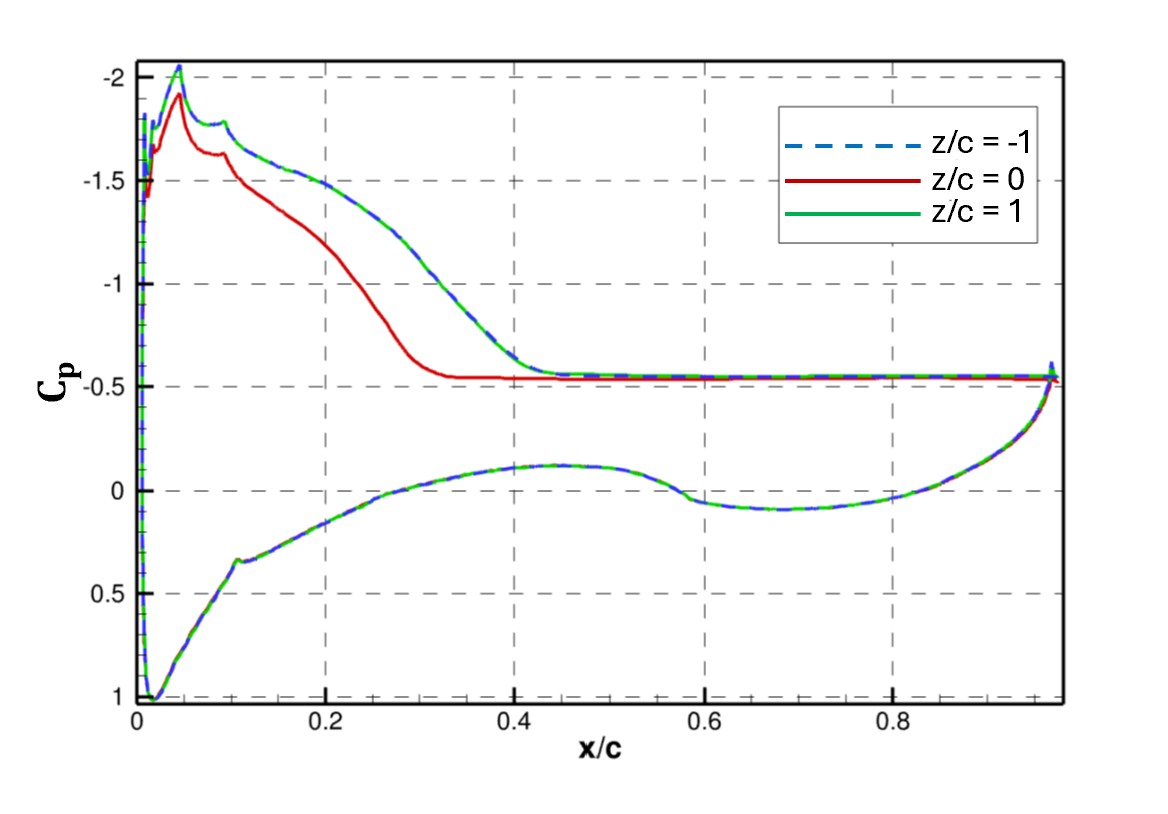}
\caption{Chordwise $C_p$ distribution at different spanwise positions.}
\label{fig:pressure distribution}
\end{figure}

To quantify these observations, Figure \ref{fig:pressure distribution} shows chordwise $C_p$ distribution curves at three spanwise positions: $z/c = 0$ (mid-span) and $z/c = \pm1$. The lift produced at mid-span (proportional to the area enclosed by the $C_p$ curve) is noticeably lower than at the other two spanwise positions. Furthermore, the $C_p$ distributions at $z/c = -1$ and $z/c = 1$ overlap almost perfectly, confirming the symmetry of the load distribution around the mid-span.

This non-uniform distribution of pressure demonstrates that in separated flows, the aerodynamic loads vary significantly along the span. Consequently, measurements or calculations at a single spanwise position would be insufficient to characterise the complete aerodynamic behavior of the airfoil. The pattern observed here, with lower loads at mid-span and higher loads at $z/c = \pm1$, is consistent with the presence of stall cells as described by \cite{Bertagnolio2005}.

This spanwise load variation has important practical implications for applications such as wind turbine blades, where it can result in uneven structural loading. A Previous study by \cite{sorensen2012computation} has shown that stall cells occur in wind turbine blade simulations with low turbulence intensity inflows. Our current study, conducted with a moderately high inflow turbulence intensity of 3\%, demonstrates that these structures persist under more realistic turbulent conditions, suggesting their relevance to practical applications.

\section{Wall friction and mean kinetic energy analysis} \label{sec:s5}

\begin{figure}
    \centering
    \includegraphics[width=1.0\linewidth]{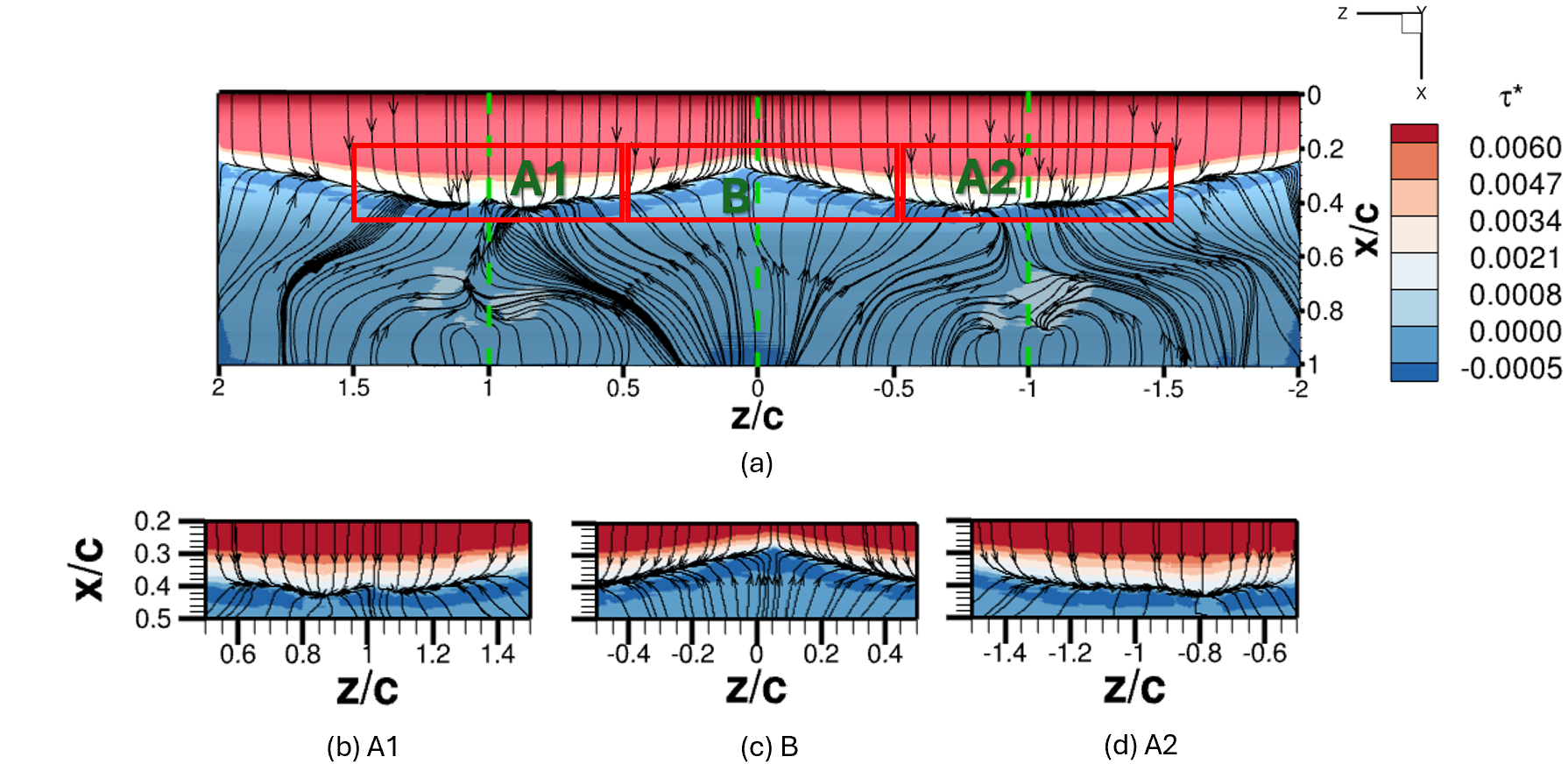}
    \caption{(a) Friction lines on the suction side of the airfoil along with the contours of the normalised $x$-wall shear stress. (b-d) Close-up view of the friction line bifurcation zone B, and convergence zones A1 and A2.}
    \label{fig:frictionline convergence and bifurcation points}
\end{figure}

To understand the flow separation patterns that generate the observed load distribution, we analyse the surface friction lines and wall shear stress on the airfoil. Figure \ref{fig:frictionline convergence and bifurcation points} (a) shows the friction lines over the surface of the airfoil along with the normalized $x$-wall shear stress distribution, $\tau^* = \tau_x/(0.5\rho U_\infty^2)$, where $\tau_x$ is the $x$-component of wall shear stress and $U_\infty$ is the freestream velocity. By definition, flow separation occurs where $\tau^*$ becomes zero.

The figure reveals that the separation point varies significantly along the span of the airfoil. Separation occurs earliest at mid-span ($z/c \approx 0$) and latest at $z/c \approx \pm1$, consistent with the pressure distribution analysis. To understand this spanwise variation, we examined the friction lines in detail. As shown in Figure \ref{fig:frictionline convergence and bifurcation points}(b-d), the friction lines remain parallel before separation but exhibit distinctly different patterns at different spanwise positions at separation.

\begin{figure}
    \centering
    \includegraphics[width=\linewidth]{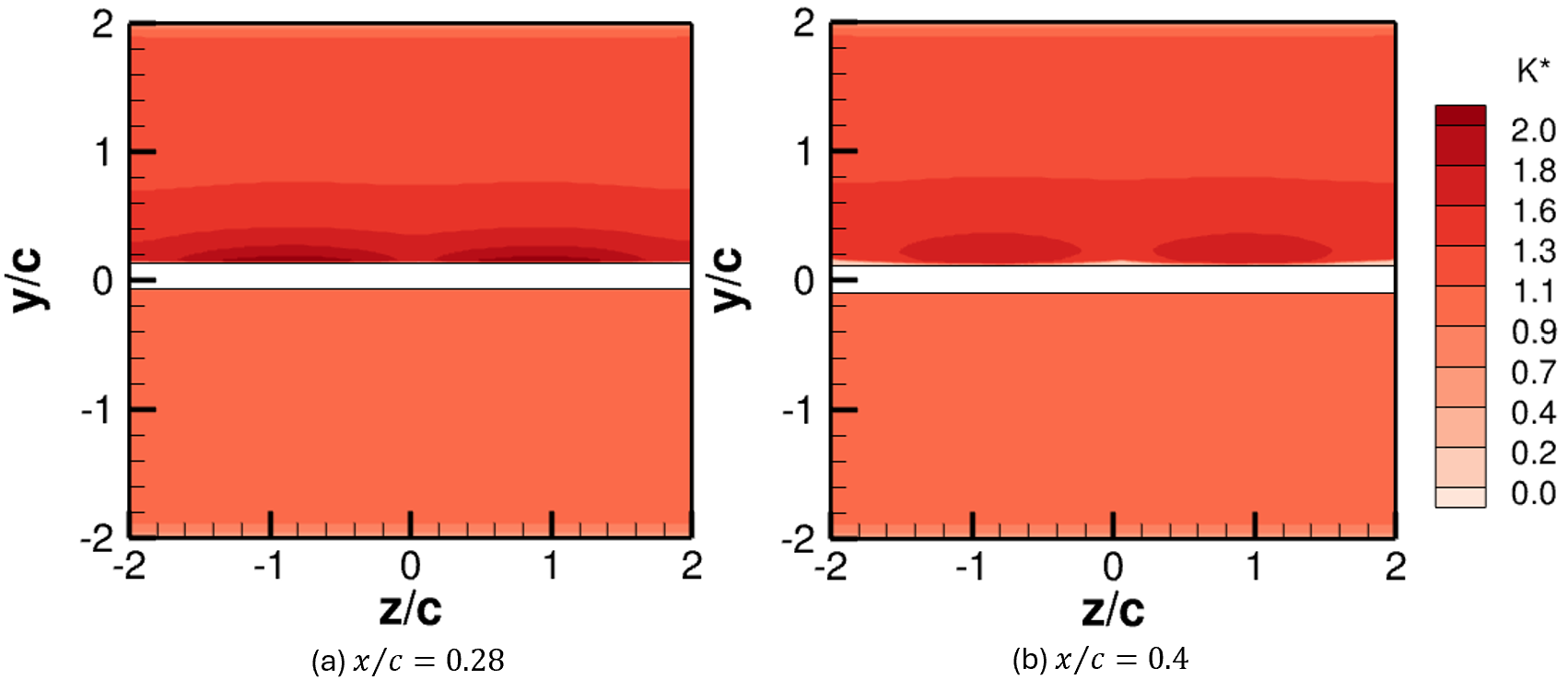}
    \caption{Contour of energy $K^*$ at (a) $x/c$ = 0.28, and (b) $x/c$ = 0.4.}
    \label{fig:energy contour plot different down stream position}
\end{figure}

At mid-span (point B in Figure \ref{fig:frictionline convergence and bifurcation points} (c)), the flow bifurcates due to spanwise instability \citep{rodriguez2011birth}. This bifurcation redirects flow toward both sides, making point B an unstable saddle point according to the classification of \cite{mazzi2020eulerian}. The bifurcation reduces flow energy near the surface at mid-span, causing earlier separation at $x/c = 0.28$. Figure \ref{fig:energy contour plot different down stream position} (a) confirms this energy reduction, showing a contour plot of normalized mean flow energy $K^* = (\sum_{i=1}^{3} U_i^2)/U_\infty^2$ at $x/c = 0.28$, with noticeably lower energy near the surface at $z/c = 0$.

In contrast, at $z/c \approx \pm1$ (points A1 and A2 in Figure \ref{fig:frictionline convergence and bifurcation points} (b), and (d)), the flow converges from both sides, creating stable nodes. These convergence zones exhibit higher flow energy near the surface, as shown in Figure \ref{fig:energy contour plot different down stream position} (b), which displays $K^*$ at $x/c = 0.4$. This higher energy enables the flow to overcome the adverse pressure gradient and separate further downstream at $x/c = 0.4$.

This analysis reveals a fundamental mechanism driving the three-dimensional organisation of separated flows: spanwise instability creates alternating regions of flow bifurcation and convergence, leading to spanwise variations in separation location, which in turn produces the observed non-uniform load distribution.

\section{Observation of stall cells} \label{sec:s6}

\begin{figure}
    \centering
    \includegraphics[width=1.0\linewidth]{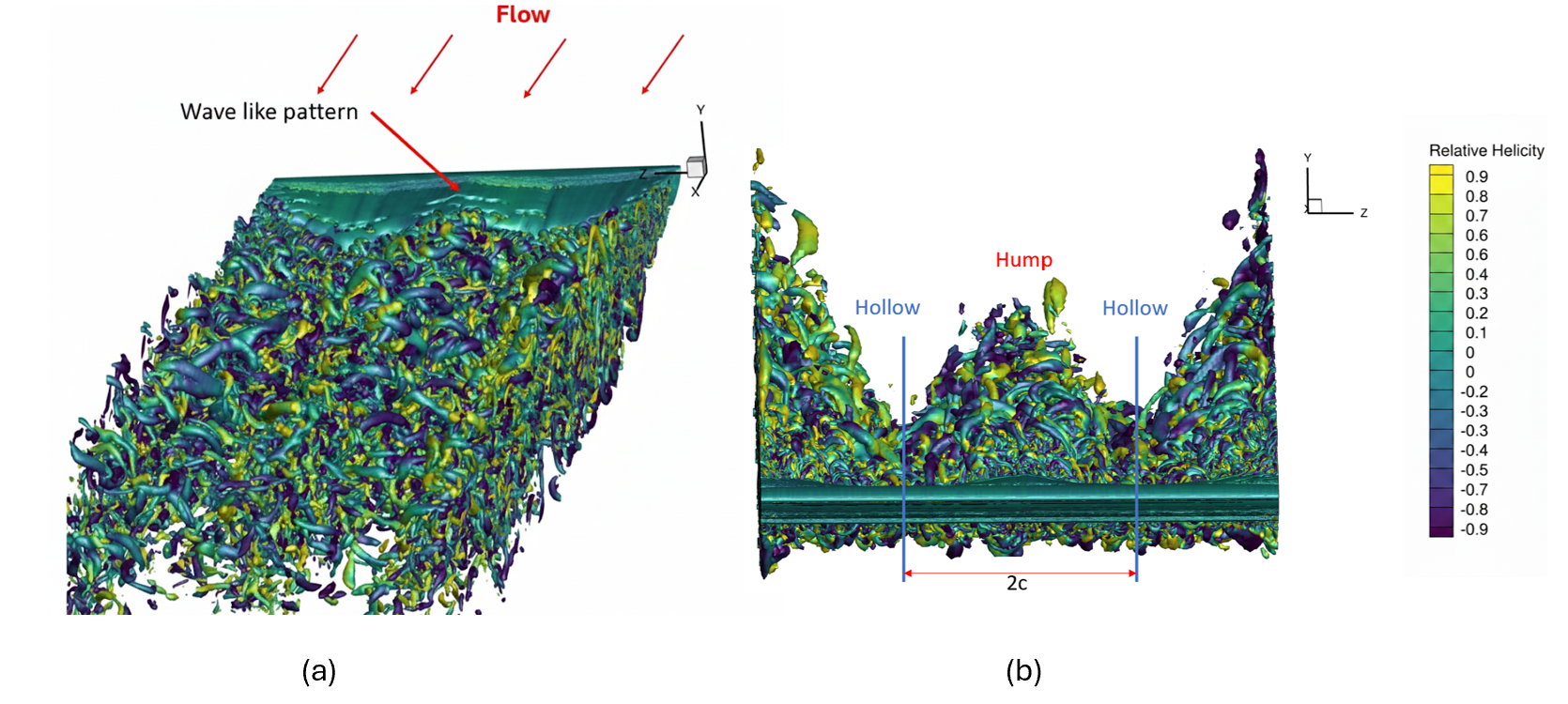}
    \caption{Instantaneous $Q = 1$ iso-surface for 14° AoA. (a) 3D view, (b) Front view}
    \label{fig:Q criterion isosurface 3D}
\end{figure}

To visualise the three-dimensional structure of the separated flow, Figure \ref{fig:Q criterion isosurface 3D} (a) shows the normalised Q-criterion iso-surface of the instantaneous flow over the airfoil for $Q = 1$, where $Q$ is defined as:

\begin{equation}
Q = \left(\frac{1}{2} \left( \|\mathbf{\Omega}\|^2 - \|\mathbf{S}\|^2 \right)\right)\frac{c^2}{U_{\infty}^2},
\end{equation}

The quantities $\|\mathbf{\Omega}\|$ and $\|\mathbf{S}\|$ denote the Frobenius norms of the vorticity tensor ($\mathbf{\Omega}$) and the strain rate tensor ($\mathbf{S}$), respectively, defined as:

\begin{equation}
    \Omega_{ij} = \frac{1}{2} \left( \frac{\partial U_i}{\partial x_j} - \frac{\partial U_j}{\partial x_i} \right).
    \label{eq:vorticity tensor}
\end{equation}

\begin{equation}
    S_{ij} = \frac{1}{2} \left( \frac{\partial U_i}{\partial x_j} + \frac{\partial U_j}{\partial x_i} \right).
    \label{eq:strain rate tensor}
\end{equation}

`


The visualisation reveals a well-separated flow with highly mixed vortical structures in the separated shear layer and wake region. A distinctive wave-like pattern is observed on the airfoil surface. Figure \ref{fig:Q criterion isosurface 3D} (b) provides a front view of this flow structure, showing more pronounced vortical structures in the mid-span region (referred to as the `hump') and near the wind tunnel wall region of the airfoil, with relatively fewer vortical structures in the intermediate regions (referred to as `hollows').


This pattern is characteristic of airfoils exhibiting trailing edge stall, where the flow does not separate uniformly along the span but forms stall cells \citep{winkelmann1980effects, schewe2001reynolds}. The humps observed in the mid-span and near-wall regions of the airfoil are consistent with the stall cell structure described by \cite{Johari2007}.

\begin{figure}
\centering
\includegraphics[width=0.5\textwidth]{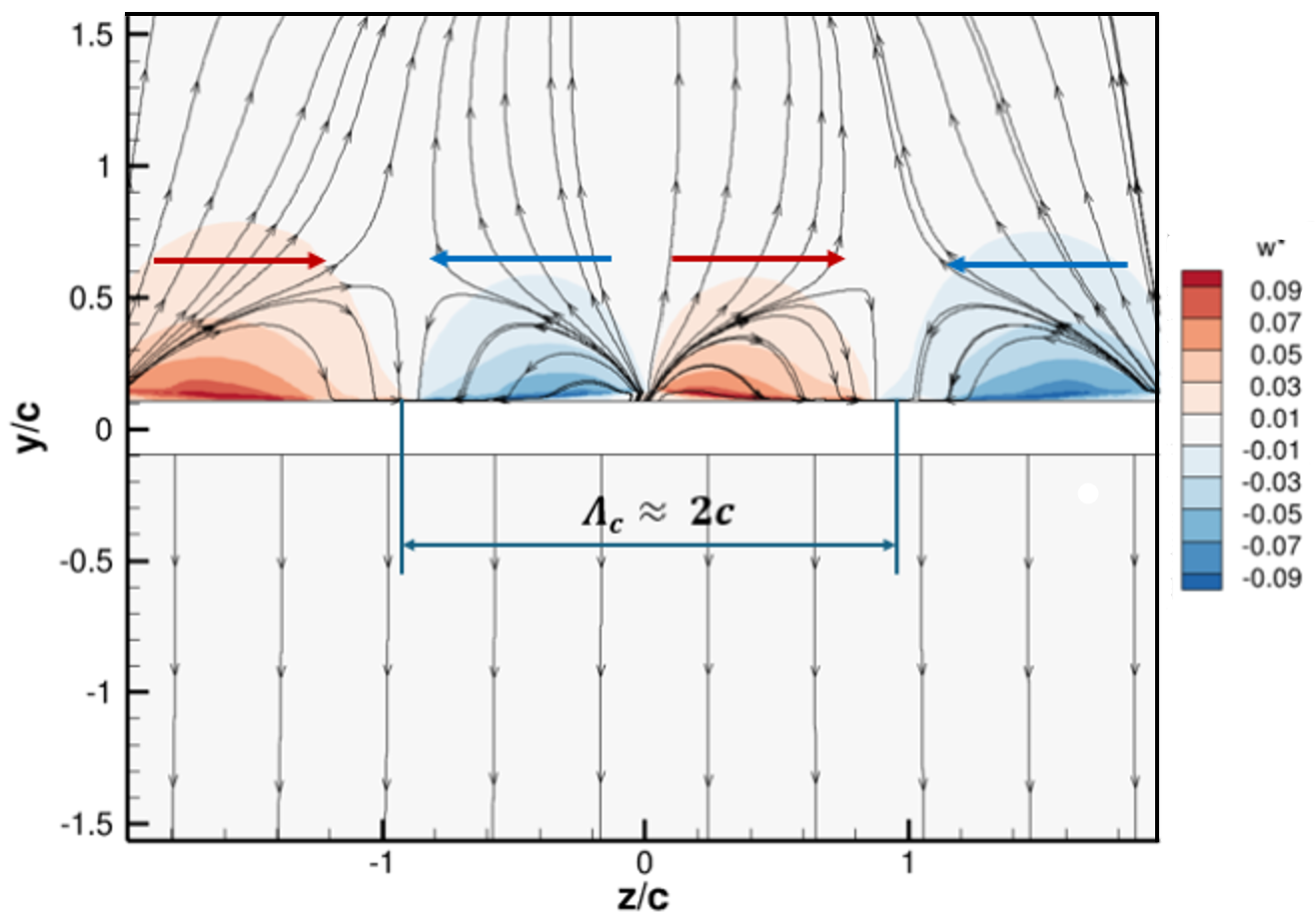}
\caption{Contour of spanwise velocity $w^*$ at $x/c = 0.4$. The red arrows indicate a positive spanwise velocity direction, while the blue arrows indicate a negative spanwise velocity direction. The black lines show in-plane streamlines.}
\label{fig:Stall cells}
\end{figure}

\begin{figure}
\centering
\includegraphics[width=0.8\textwidth]{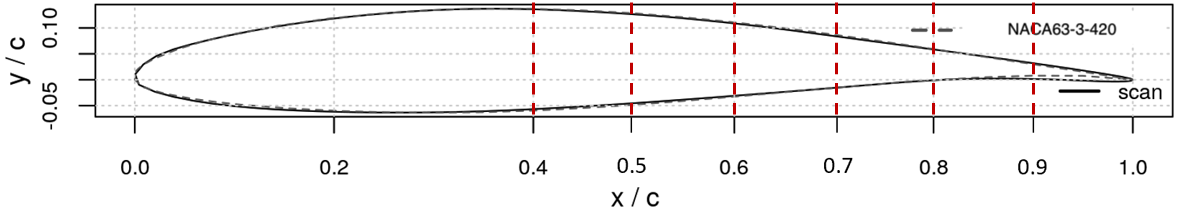}
\caption{Location of different downstream positions (red dashed lines) on the airfoil}
\label{fig:different downstream positions on airfoil}
\end{figure}

To confirm the presence of stall cells, Figure \ref{fig:Stall cells} presents a contour plot of the normalized spanwise velocity ($w^* = W/U_{\infty}$) at $x/c = 0.4$, superimposed with in-plane streamlines. Figure \ref{fig:different downstream positions on airfoil} shows the locations of the various downstream positions referenced throughout the analysis.. The near-red regions indicate spanwise velocity in the positive z-direction, while the blue regions indicate spanwise velocity in the negative z-direction. These pockets of relatively high spanwise velocities ($|w^*| = 0.5$ on average) represent the core regions of stall cells.

One stall cell is defined as the pair of positive and negative spanwise velocity contours. The stall cells observed at this angle of attack are approximately the twice the size of the airfoil chord, which agrees well with the findings of \cite{schewe2001reynolds}, where measurements were taken at 12° angle of attack. Size of one stall cell forms one wavelength ($\Lambda_c$) \citep{manni2016numerical}, resulting in $\Lambda_c \approx 2c$ between $z/c = 1$ and $z/c = -1$ in the present study. This result is consistent with similar experimental studies on thick airfoils \citep{schewe2001reynolds, yon1998study, broeren2001spanwise}, which found $\Lambda_c \approx 2c$ for an aspect ratio of 4 with comparable Reynolds numbers. \\

In the next section the origin of these alternating positive and negative spanwise velocity contours is discussed.

\section{Mean flow field analysis} \label{sec:s7}

\subsection{The separated shear layer}

\begin{figure}
    \centering
    \includegraphics[width=1.0\linewidth]{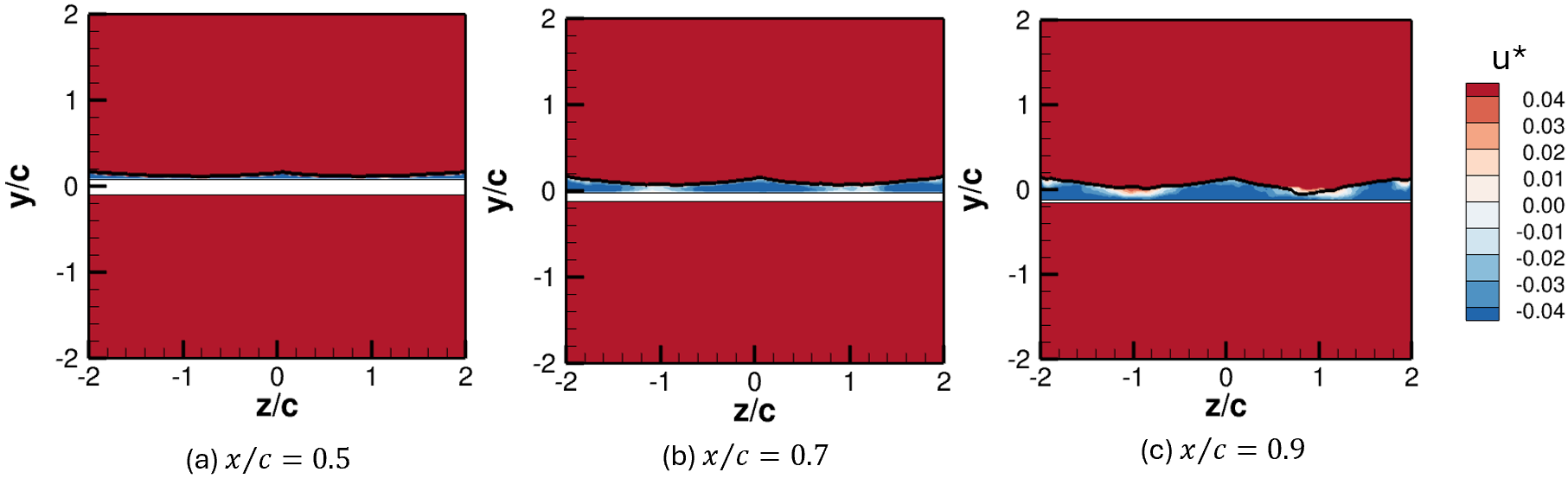}
    \caption{Contour plots of normalised downstream velocity $u^*$ at (a) $x/c$ = 0.5, (b) $x/c$ = 0.7, and (c) $x/c$ = 0.9. The black curve represents $u^* = 0$ iso-curve.}
    \label{fig:u contour plot different down stream position}
\end{figure}

Flow separation on the airfoil creates a suction zone that induces a reverse flow in the upstream direction. Figure \ref{fig:u contour plot different down stream position} shows contours of the normalized streamwise velocity component $u^* = U/U_{\infty}$ at $x/c = 0.5, 0.7,$ and $0.9$, respectively. In these contour plots, the $u^* = 0$ iso-curve (black line) divides the flow into two distinct regions: the reverse flow region between the airfoil surface and the $u^* = 0$ curve (predominantly blue), and the freestream flow region above the $u^* = 0$ curve (predominantly red).

\begin{figure}
\centering
\includegraphics[width=\textwidth]{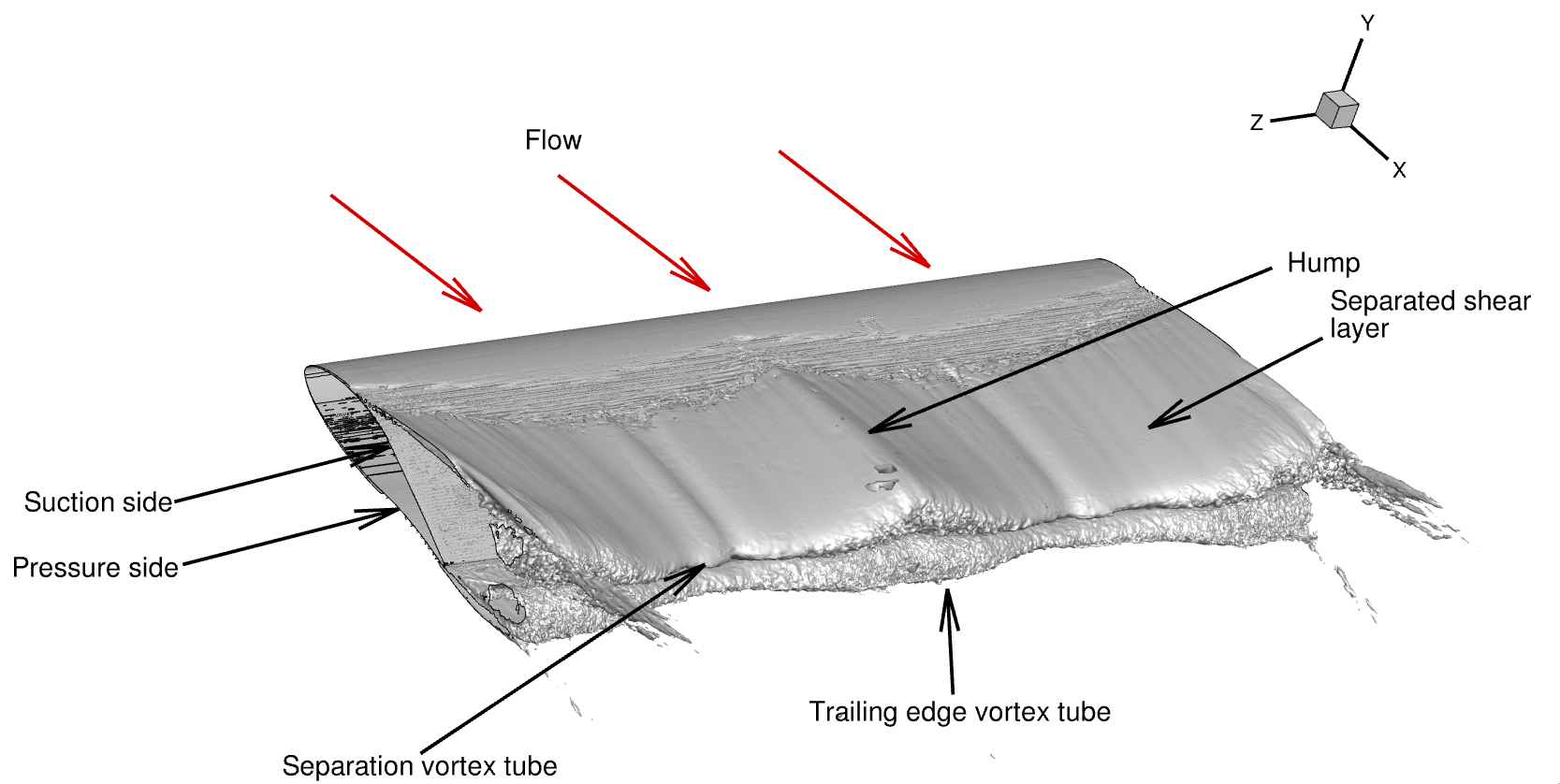}
\caption{Averaged $Q=1$ iso-surface showing the shear layer, its wavelike form, and the hump seen on this layer.}
\label{fig: Averaged Q =1 criterion}
\end{figure}

The area traced by the $u^* = 0$ iso-surface defines the shear layer region, which gives rise to the separation vortex sheet and tube. This separated shear layer can be visualised using the iso-surface of $Q = 1$, as shown in Figure \ref{fig: Averaged Q =1 criterion}. Notably, a wave-like pattern appears on the averaged $Q$ iso-surface, representing the vortex sheet, which terminates in a similarly wave-like bend on the separation vortex tube.

\begin{figure}
\centering
\subfloat[$v^*$]{\includegraphics[width=0.48\textwidth]{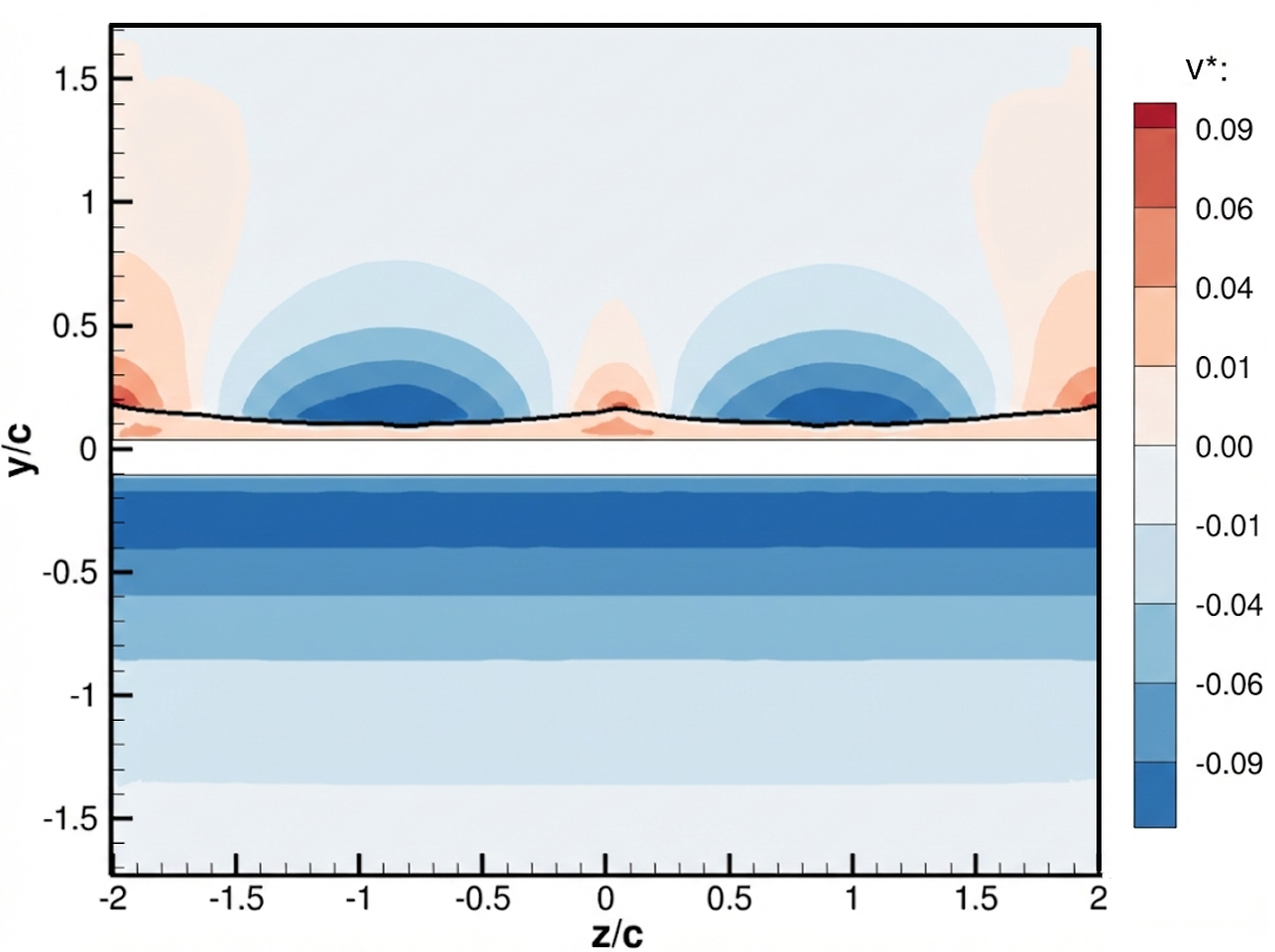}
\label{fig:70 percent v* contour}}
\hfill
\subfloat[$w^*$]{\includegraphics[width=0.48\textwidth]{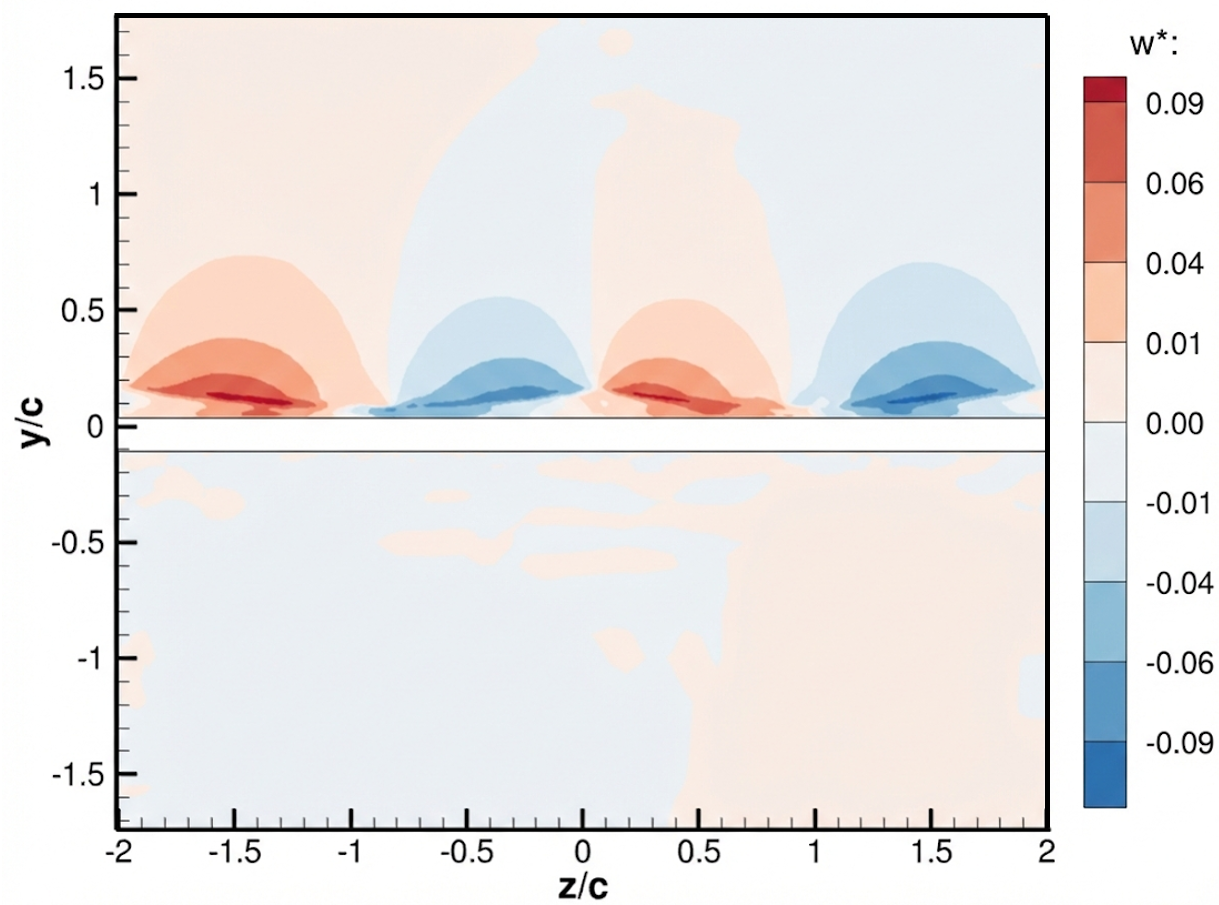}
\label{fig:80 percent v* contour}}
\caption{Contour plot of normalised (a) $y$-velocity $v^*$ and (b) $z$-velocity $w^*$ at $x/c = 0.6$. The black curve in figure (a) represents cross-sections at $x/c = 0.6$ of the averaged $Q = 1$ iso-surface, indicating the shear layer.}
\label{fig:v* contour at different doanstream condition}
\end{figure}

The three-dimensional nature of the flow is further evidenced by examining the vertical velocity component $v^* = V/U_{\infty}$ and spanwise velocity component $w^* = W/U_{\infty}$, shown in Figure \ref{fig:v* contour at different doanstream condition} at $x/c = 0.6$. Both components exhibit alternating patterns in the spanwise direction and have magnitudes comparable to each other, approximately 10\% of the longitudinal velocity $u^*$. This three-dimensional organisation necessitates analysis of the vorticity components to understand the underlying mechanisms.

\subsection{Vorticity analysis}

Examination of the different vorticity components reveals that spanwise vorticity ($\Omega_z$) dominates over streamwise vorticity ($\Omega_x$) and vertical vorticity ($\Omega_y$) in the separation region, primarily due to the strong shear generated by flow separation. However, all three components interact to create the observed three-dimensional flow organisation.

\begin{figure}
\centering
\includegraphics[width=0.8\textwidth]{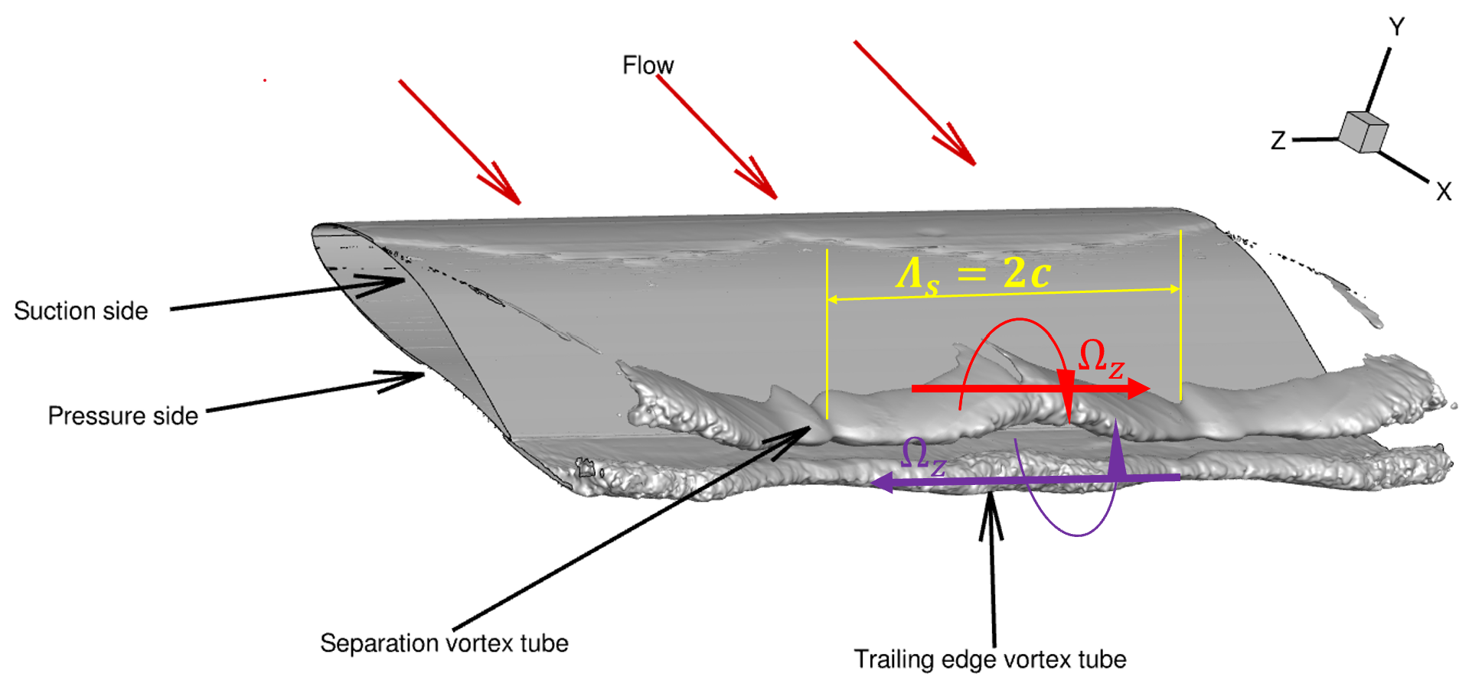}
\caption{Averaged $Q=7$ iso-surface showing the separation and trailing edge vortex tubes.}
\label{fig: Averaged Q criterion}
\end{figure}

\begin{figure}
\centering
\includegraphics[width=0.6\textwidth]{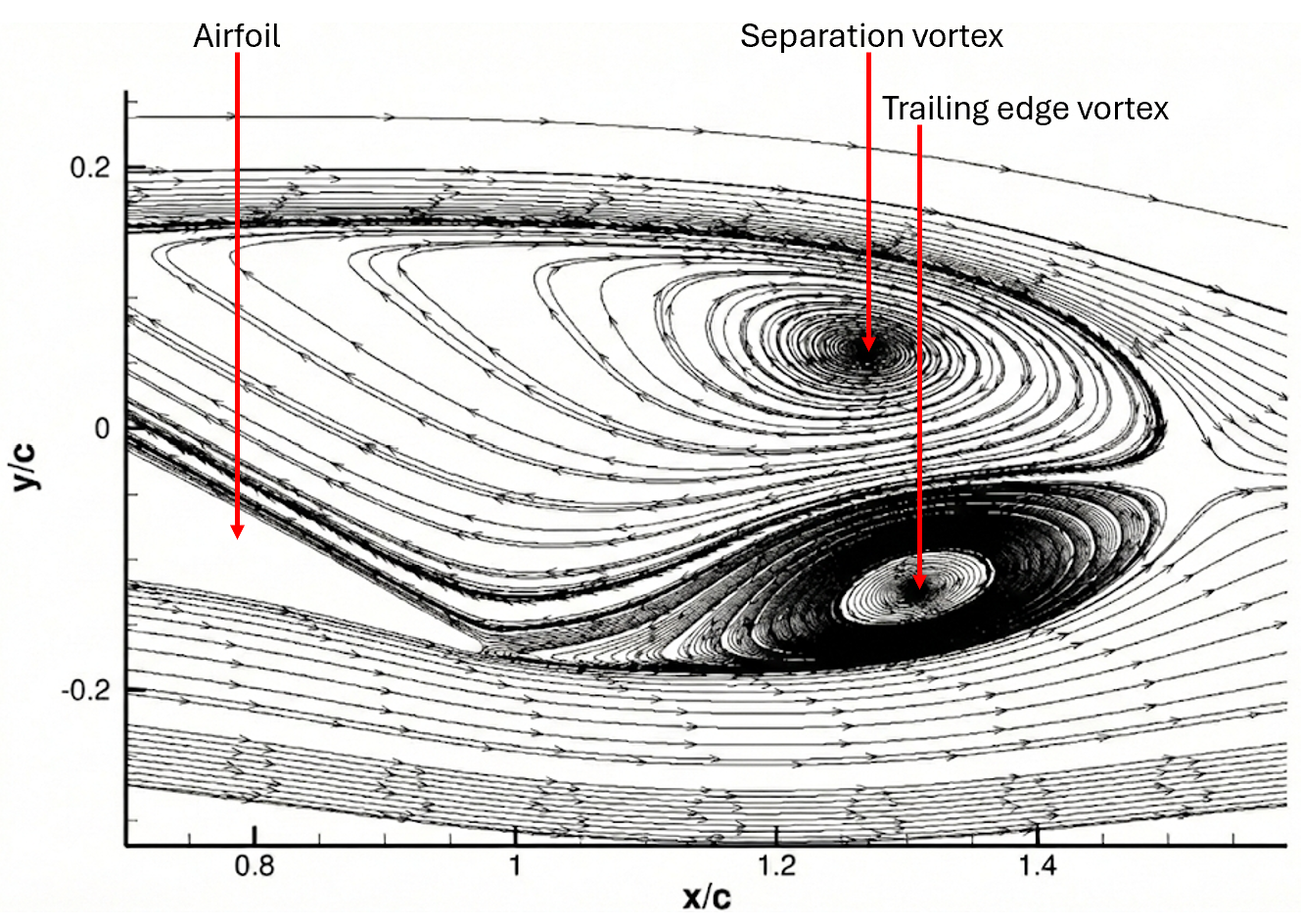}
\caption{Streamlines plotted at the $z/c = 0$ plane showing the separation vortex and trailing vortex cores.}
\label{fig:streamline z/c = 0}
\end{figure}

\begin{figure}
    \centering
    \includegraphics[width=1\linewidth]{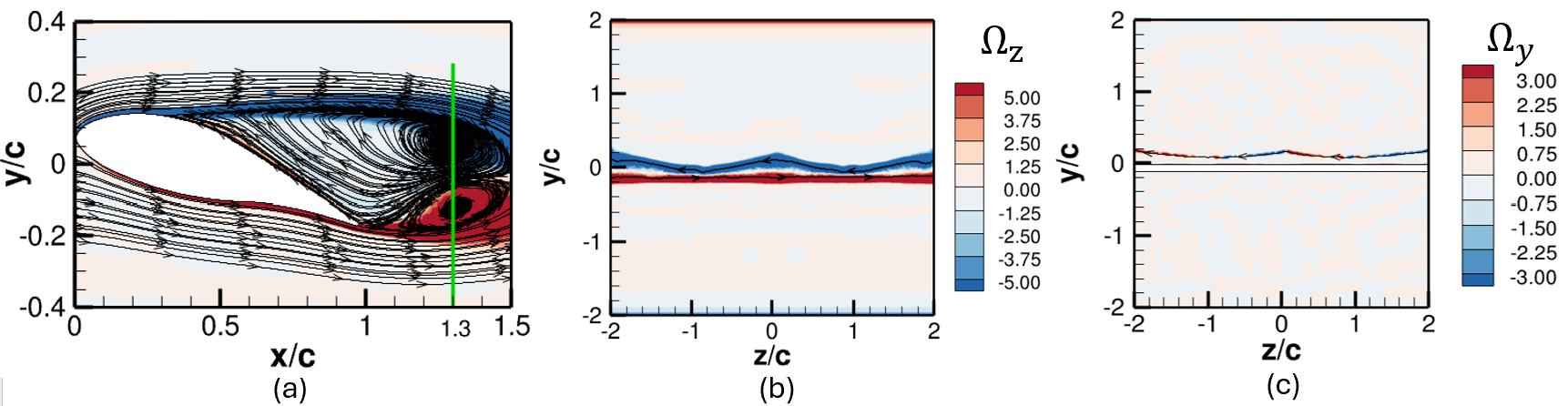}
    \caption{visualisation of the vorticity field. (a) $\Omega_z$ field at the symmetry plane ($z/c = 0$) with in-plane streamlines, where the green line at $x/c = 1.3$ identifies the vortex core center. (b) Transverse view of $\Omega_z$ at the core center ($x/c = 1.3$) and (c) $\Omega_y$ upstream at $x/c = 0.7$, both plotted with in-plane vortex lines.}
    \label{fig:Wz vorticity field with streamlines at palne x/c = 1.3}
\end{figure}

Figure \ref{fig: Averaged Q criterion} shows the iso-surface at $Q = 7$, revealing both the separated vortex and the mean trailing edge vortex tubes. Figure \ref{fig:streamline z/c = 0} highlights these features at $z/c = 0$ using in-plane streamlines. These vortex tubes form a pair of counter-rotating vortices, which \cite{crow1970stability} found to be inherently unstable - a finding later confirmed by \cite{klein1995simplified} and \cite{fabre2002instabilite}. Figure \ref{fig:Wz vorticity field with streamlines at palne x/c = 1.3} (a) provides another view of the $\Omega_z$ field, showing the centers of the vortex cores around $x/c = 1.3$.

This Crow-type instability causes the vortex tubes to bend in a wave-like pattern when influenced by a parallel counter-rotating vortex tube. Figure \ref{fig:Wz vorticity field with streamlines at palne x/c = 1.3} (b) clearly shows this bending in a contour plot of normalized spanwise vorticity ($\Omega_z$) together with in-plane vortex lines at $x/c = 1.3$. The in-plane vortex lines are defined by:

\begin{equation}
    \frac{\Omega_y}{dy} = \frac{\Omega_z}{dz}.
    \label{eq:definition of in-plane vortex lines}
\end{equation}

The wavelength of the bend in the vortex tubes ($\Lambda_s$) is approximately $2c$, which equals the stall cell wavelength $\Lambda_c$. Importantly, this separation vortex tube is not an independent structure but forms the end of the vortex sheet. The bending of the vortex tubes is therefore correlated with a bend in the vortex sheet, appearing as a hump in the vortex sheet and explaining the wave-like pattern seen in the shear layer in the spanwise direction.

This bending of the vortex tube and vortex sheet induces vertical vorticity ($\Omega_y$) (see Figure \ref{fig:Wz vorticity field with streamlines at palne x/c = 1.3} (c), defined as:

\begin{equation}
    \Omega_y = \left(\frac{\partial U}{\partial z}-\frac{\partial W}{\partial x}\right)\frac{c}{U_{\infty}}.
    \label{eq:definition of the y vorticity}
\end{equation}

\begin{figure}
    \centering
    \includegraphics[width=\linewidth]{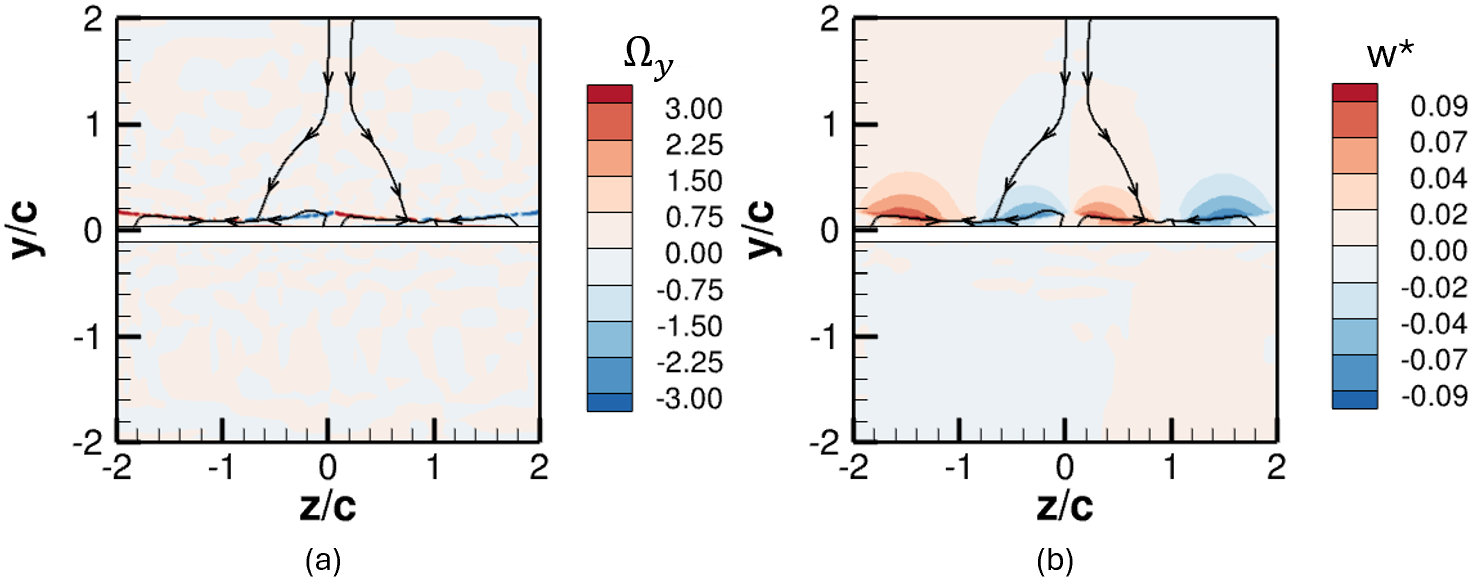}
    \caption{Contour plots from DDES at $x/c = 0.6$: (a) normalised vertical vorticity $\Omega_y$ showing alternating positive and negative regions; (b) normalised spanwise velocity $w^*$ exhibiting the characteristic stall cell pattern. The black curves are showing the in-plane streamlines.}
    \label{fig:omega_y_w}
\end{figure}

The induction of $\Omega_y$ generates spanwise velocity $w^*$, contributing significantly to the three-dimensional nature of the flow separation. Figure \ref{fig:omega_y_w} (a) shows a contour plot of $\Omega_y$ at $x/c = 0.6$. The alternating positive and negative regions of $\Omega_y$ are seen to be aligned with the positive and negative turning streamlines. This explain the corresponding alternating positive and negative spanwise velocities, for example, seen in the Figure \ref{fig:omega_y_w} (b).

\begin{figure}
\centering
\includegraphics[width=0.6\textwidth]{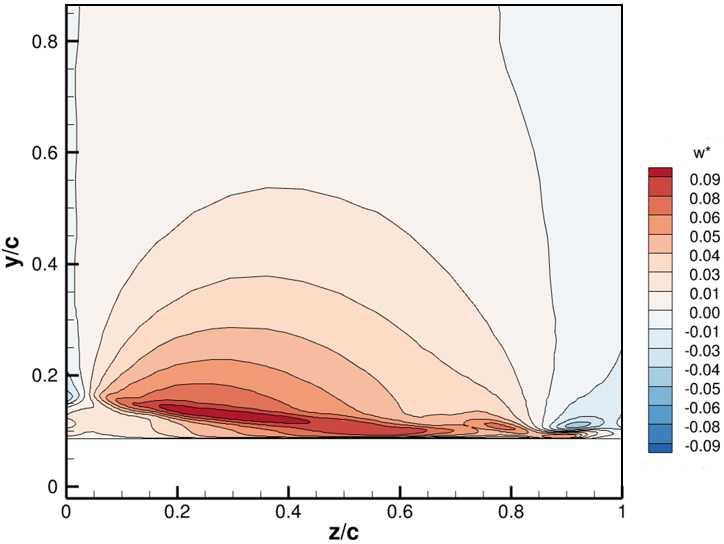}
\caption{$w^*$ velocity structure contour plot showing one cell between $z/c = 0$ and $z/c = 1$ at $x/c = 0.5$ position.}
\label{fig: w velocity cell close view}
\end{figure}

\begin{figure}
\centering
\includegraphics[width=0.8\textwidth]{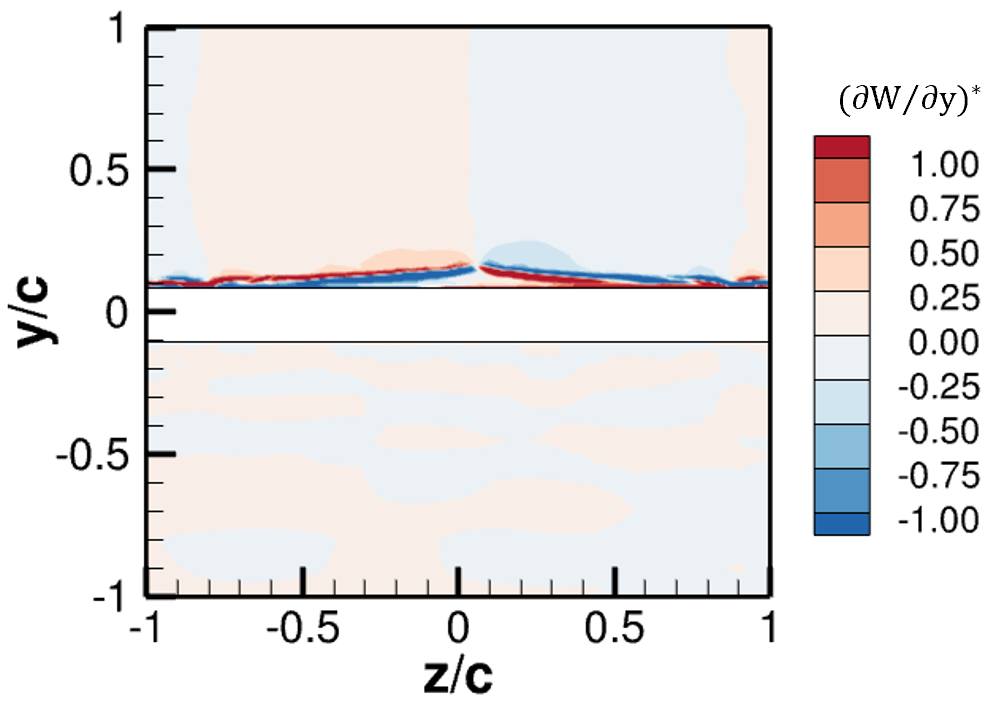}
\caption{Contour plot of normalised vertical gradient of $w^*$, i.e., $\left(\frac{\partial W}{\partial y}\right)^*$ between $z/c = 1$ and $z/c = -1$ at $x/c = 0.5$. Note that this is an enlarged view with $-1\leq y/c\leq 1$ and $-1\leq z/c\leq 1$.}
\label{fig: w velocity gradient with respect to y}
\end{figure}

Examining the spanwise velocity structures in more detail, Figure \ref{fig: w velocity cell close view} shows a close-up view of the $w^*$ cell between $z/c = 0$ and $z/c = -1$ at $x/c = 0.5$. The vertical gradient of this spanwise velocity, $\left(\frac{\partial W}{\partial y}\right)^* = \left(\frac{\partial W}{\partial y}\right) \frac{c}{U_{\infty}}$, is shown in Figure \ref{fig: w velocity gradient with respect to y}. This gradient forms a significant component of the streamwise vorticity ($\Omega_x$), defined as:

\begin{equation}
    \Omega_x = \left(\frac{\partial W}{\partial y}-\frac{\partial V}{\partial z}\right)\frac{c}{U_{\infty}}.
    \label{eq:definition of the x vorticity}
\end{equation}

\begin{figure}
\centering
\includegraphics[width=1\textwidth]{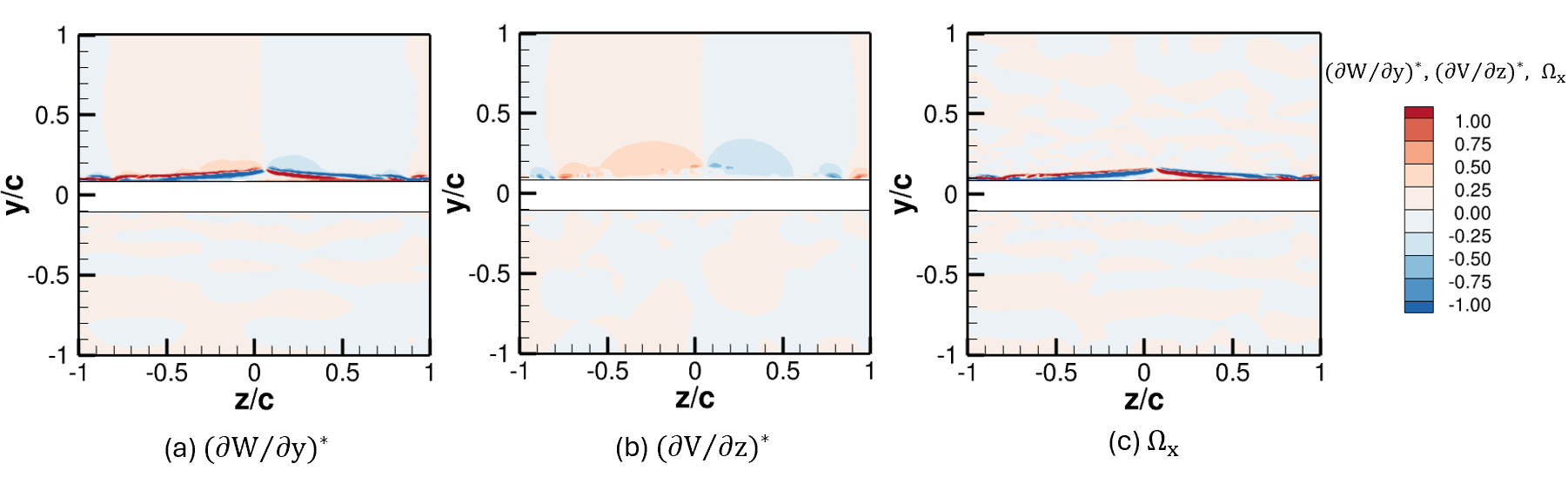}
\caption{Contour plot presenting different components of the $x$-vorticity equation (\ref{eq:definition of the x vorticity}) at $x/c = 0.5$. Note that this is an enlarged view with $-1\leq y/c\leq 1$ and $-1\leq z/c\leq 1$.}
\label{fig: contour plot representation of the definition of the x vorticity}
\end{figure}

Figure \ref{fig: contour plot representation of the definition of the x vorticity} presents the contour plot of different components of equation (\ref{eq:definition of the x vorticity}) at $x/c = 0.5$, confirming that $\Omega_x$ is predominantly influenced by the vertical gradient of the spanwise velocity $\left(\frac{\partial W}{\partial y}\right)^*$, as the patterns of $\Omega_x$ and $\left(\frac{\partial W}{\partial y}\right)^*$ are nearly identical.

\begin{figure}
\centering
\includegraphics[width=\textwidth]{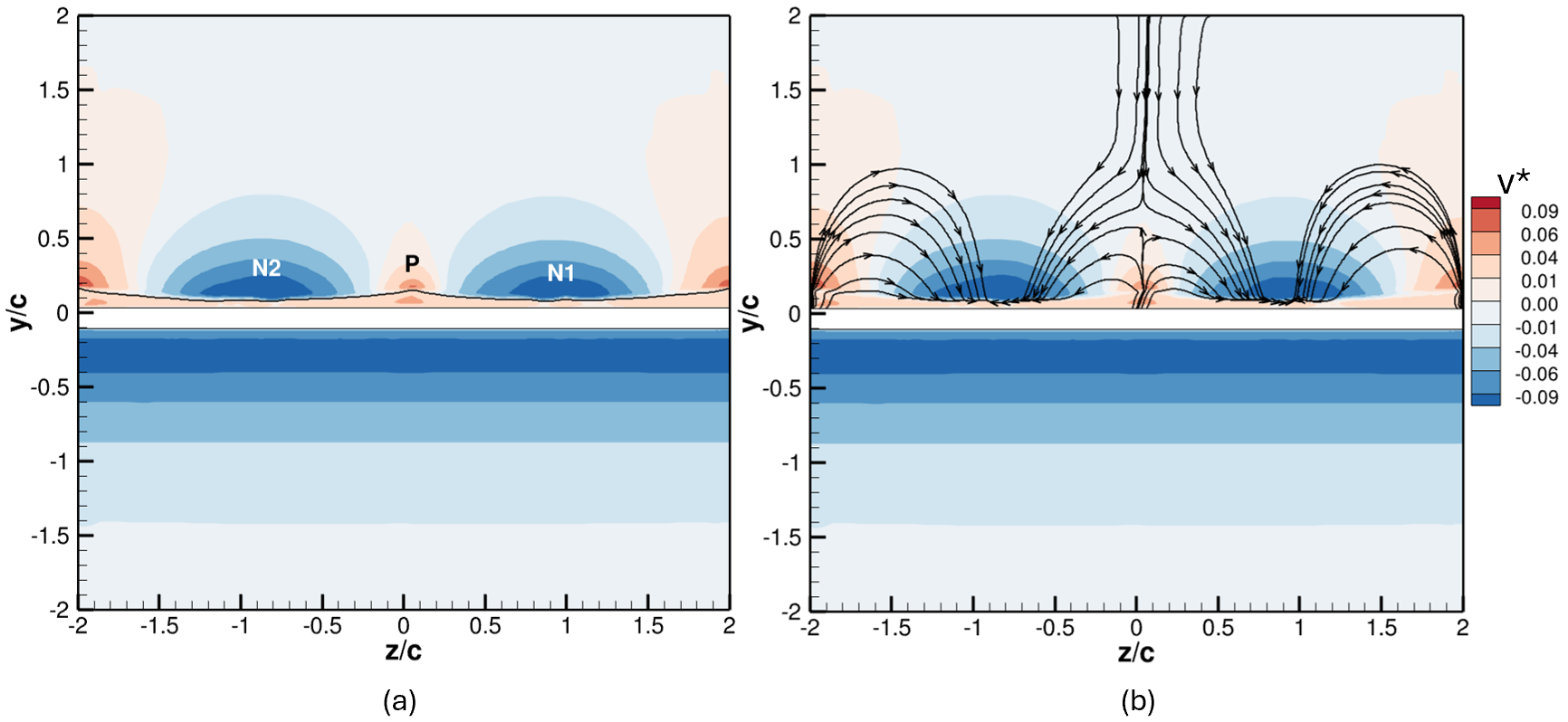}
\caption{ Contour plot representation of $v^*$ at $x/c = 0.6$. (a) N1 and N2 denote zones where $v^*$ is negative, while P denotes the region where $v^*$ is positive. The black curve represents the shear layer (cross-section of the averaged $Q = 1$ iso-surface). (b) Representation of the flow field including in-plane streamlines.}
\label{fig: contour plot representation of the definition of the y velocity}
\end{figure}

The vertical velocity field $v^*$ shows a complex organisation resulting from the interaction of $\Omega_x$ and $\Omega_z$. Figure \ref{fig: contour plot representation of the definition of the y velocity} (a) shows the $v^*$ contours at $x/c = 0.6$, with three distinct zones: a positive $v^*$ zone (P) centered around $z/c = 0$, and two negative $v^*$ zones (N1 and N2) centered around $z/c = 1$ and $z/c = -1$, respectively. Figure \ref{fig: contour plot representation of the definition of the y velocity} (b) shows this same field with in-plane streamlines, revealing that the streamlines are attracted to the $Q = 1$ iso-surface in zones N1 and N2 due to flow entrainment caused by the interaction of vortices in the shear layer and the outer flow.

\begin{figure}
\centering
\includegraphics[width=0.6\textwidth]{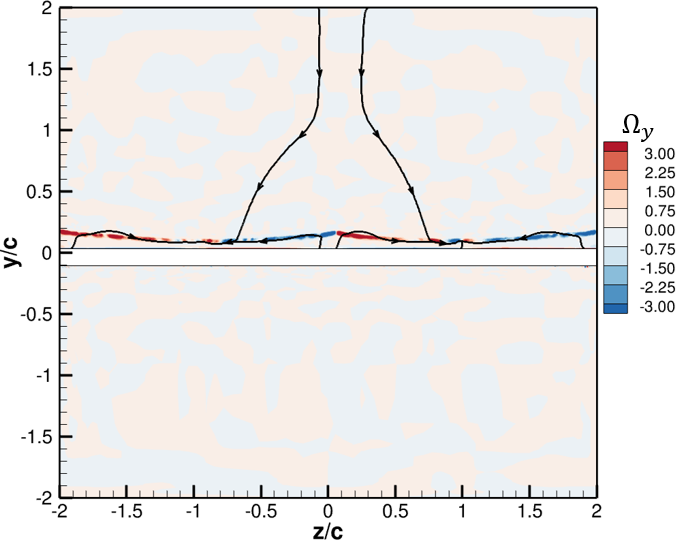}
\caption{Contour plot of $\Omega_y$ showing the left and right turning of in-plane streamlines at $x/c = 0.6$. Only a few in-plane streamlines are drawn for clarity.}
\label{fig: left right turning od streamlines}
\end{figure}

The direction of streamline turning is determined by the sign of $\Omega_y$, as illustrated in Figure \ref{fig: left right turning od streamlines}: where $\Omega_y$ is positive, the flow turns toward the $+z$ direction, and where $\Omega_y$ is negative, the flow turns toward the $-z$ direction. This alternating pattern of $\Omega_y$ thus drives the alternating spanwise flow that characterises stall cells.

\begin{figure}
    \centering
    \includegraphics[width=\linewidth]{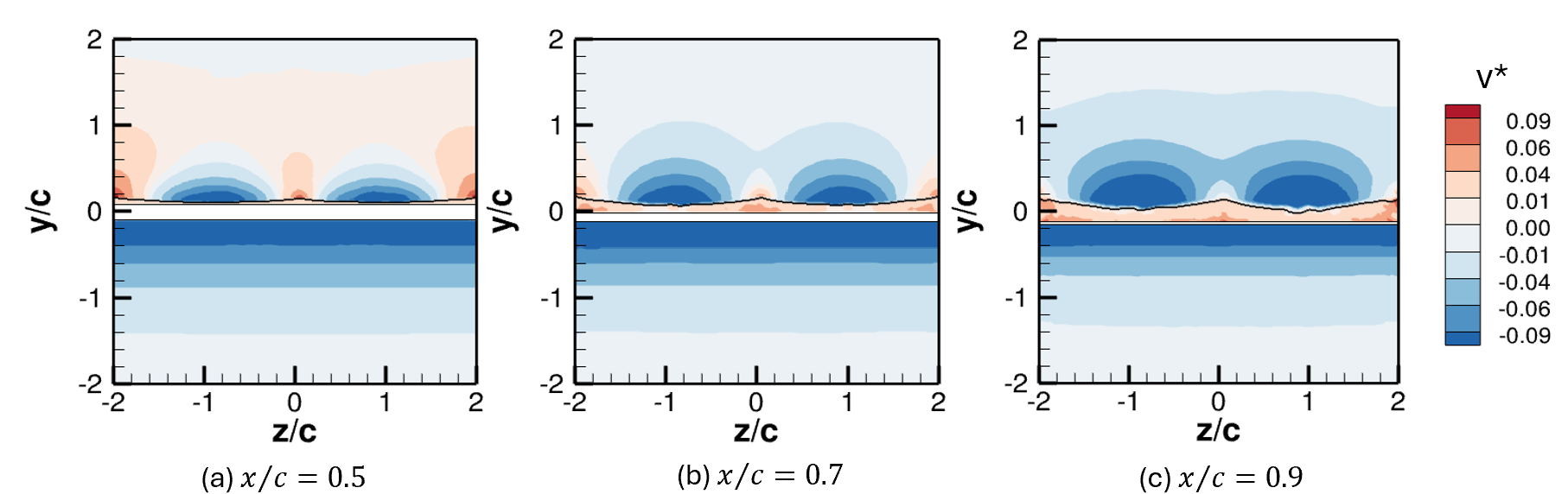}
    \caption{Contour plot of normalised $y$-velocity $v^*$ at (a) $x/c$ = 0.5, (b) $x/c$ = 0.7, and (c) $x/c$ = 0.9 positions. The black curve represents cross-sections at respective downstream positions of the averaged $Q = 1$ iso-surface, indicating the shear layer.}
    \label{fig:v* contour at different downstream condition}
\end{figure}

As the flow progresses downstream, the P zone gradually diminishes while the N1 and N2 zones intensify, as shown in Figure \ref{fig:v* contour at different downstream condition}, which displays $v^*$ contours at $x/c = 0.5, 0.7,$ and $0.9$. Between the $Q = 1$ iso-surface and the airfoil suction side, $v^*$ consistently remains positive due to the upwelling effect of the separation vortex in the separation zone.

\subsection{Streamwise evolution of the vortex core} \label{sec:Streamwise Evolution of the Vortex Core}

\begin{figure}
    \centering
    \includegraphics[width=\linewidth]{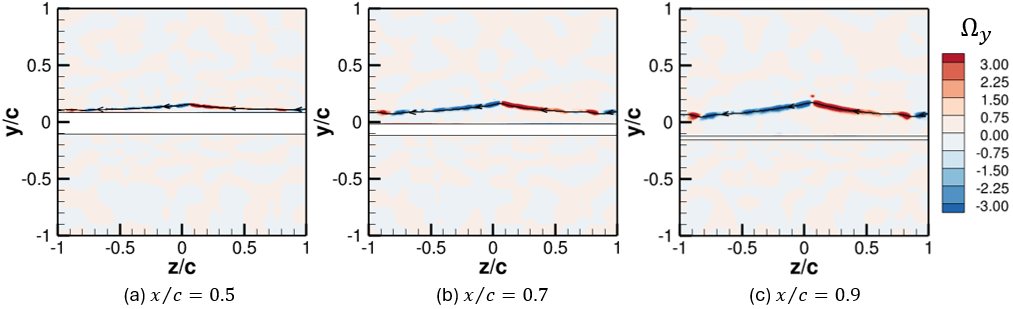}
    \caption{Contour plots of the normalised vertical vorticity $\Omega_y$ at (a) $x/c$ = 0.5, (b) $x/c$ = 0.7, and (c) $x/c$ = 0.9 along with the core vortex line. Note that this is an enlarged view with $-1\leq y/c\leq 1$ and $-1\leq z/c\leq 1$.}
    \label{fig:vertical vorticity contour plot different down stream position}
\end{figure}

The downstream evolution of the vorticity components, particularly $\Omega_x$ and $\Omega_y$, reveals interesting patterns that further elucidate the three-dimensional organisation of the flow. Figure \ref{fig:vertical vorticity contour plot different down stream position} shows contour plots of $\Omega_y$ at downstream positions $x/c = 0.5, 0.7,$ and $0.9$.  For visual clarity, zoomed versions of these contour plots within the region $-1 \leq y/c \leq 1$ and $-1 \leq z/c \leq 1$ are shown. The highest values of $\Omega_y$ occur in the separated shear layer region, and notably, the magnitude of $\Omega_y$ remains relatively constant as the flow progresses downstream.

\begin{figure}
    \centering
    \includegraphics[width=\linewidth]{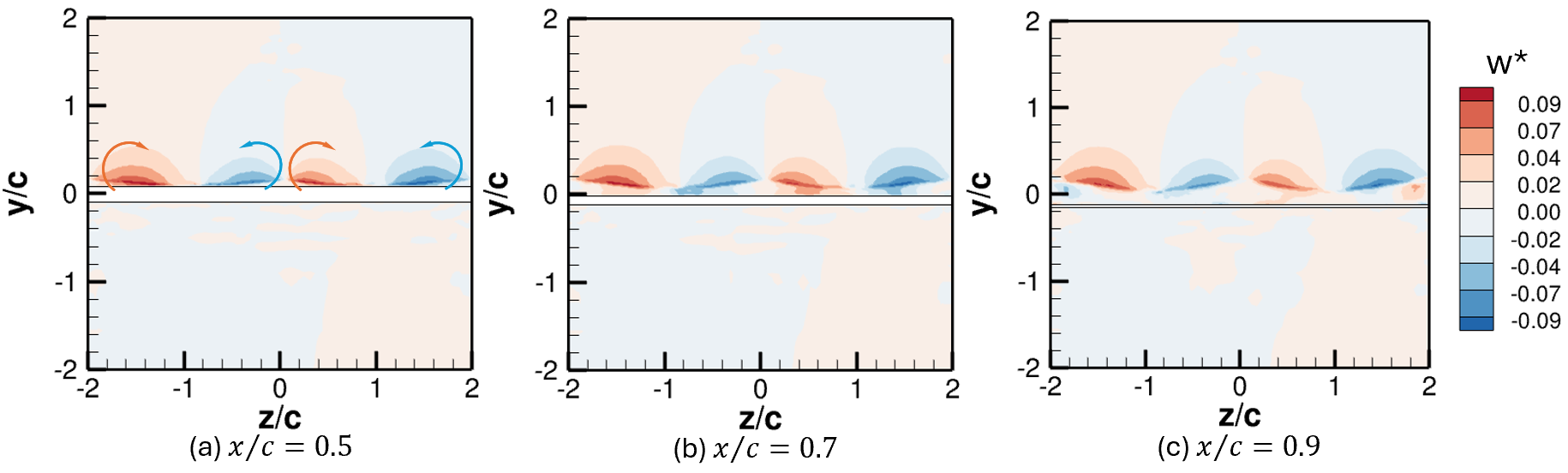}
    \caption{Contour of normalised spanwise velocity at (a) $x/c$ = 0.5, (b) $x/c$ = 0.7, and (c) $x/c$ = 0.9 positions. The curved arrows in the figure show the direction of rotation of the spanwise velocity structures.}
    \label{fig:w contour plot different down stream position}
\end{figure}

\begin{figure}
\centering
\includegraphics[width=0.6\textwidth]{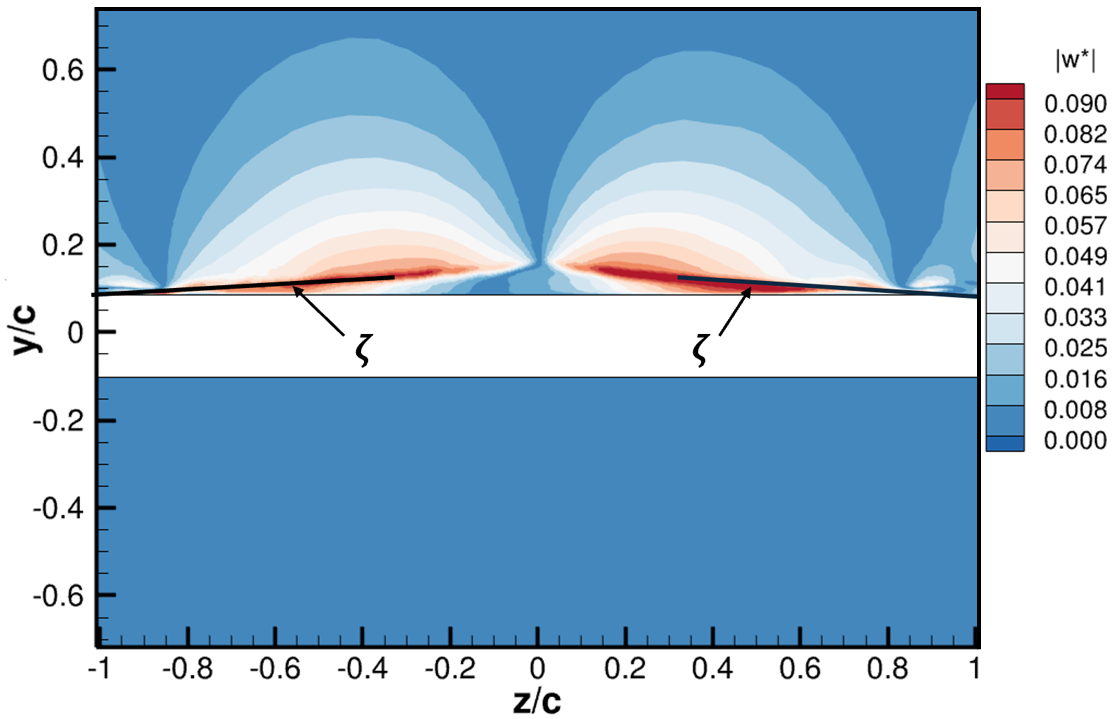}
\caption{Contour plot showing the rotation angle $\zeta$ of the $w^*$ velocity stall cell structures.}
\label{fig: Rotation angle of stall cells}
\end{figure}

The spanwise velocity structures, shown in Figure \ref{fig:w contour plot different down stream position} at various downstream positions, exhibit a remarkable rotational behavior. With increasing downstream distance, the contours of spanwise velocity rotate approximately around $z/c = \pm1$, as indicated by the red and blue arrows in Figure \ref{fig:w contour plot different down stream position} (a). By identifying the maximum values of the iso-contours of $|w^*|$ and drawing lines connecting these maxima with $z/c = \pm1$, we can measure the rotation angle $\zeta$ (Figure \ref{fig: Rotation angle of stall cells}).

\begin{figure}
\centering
\includegraphics[width=0.6\textwidth]{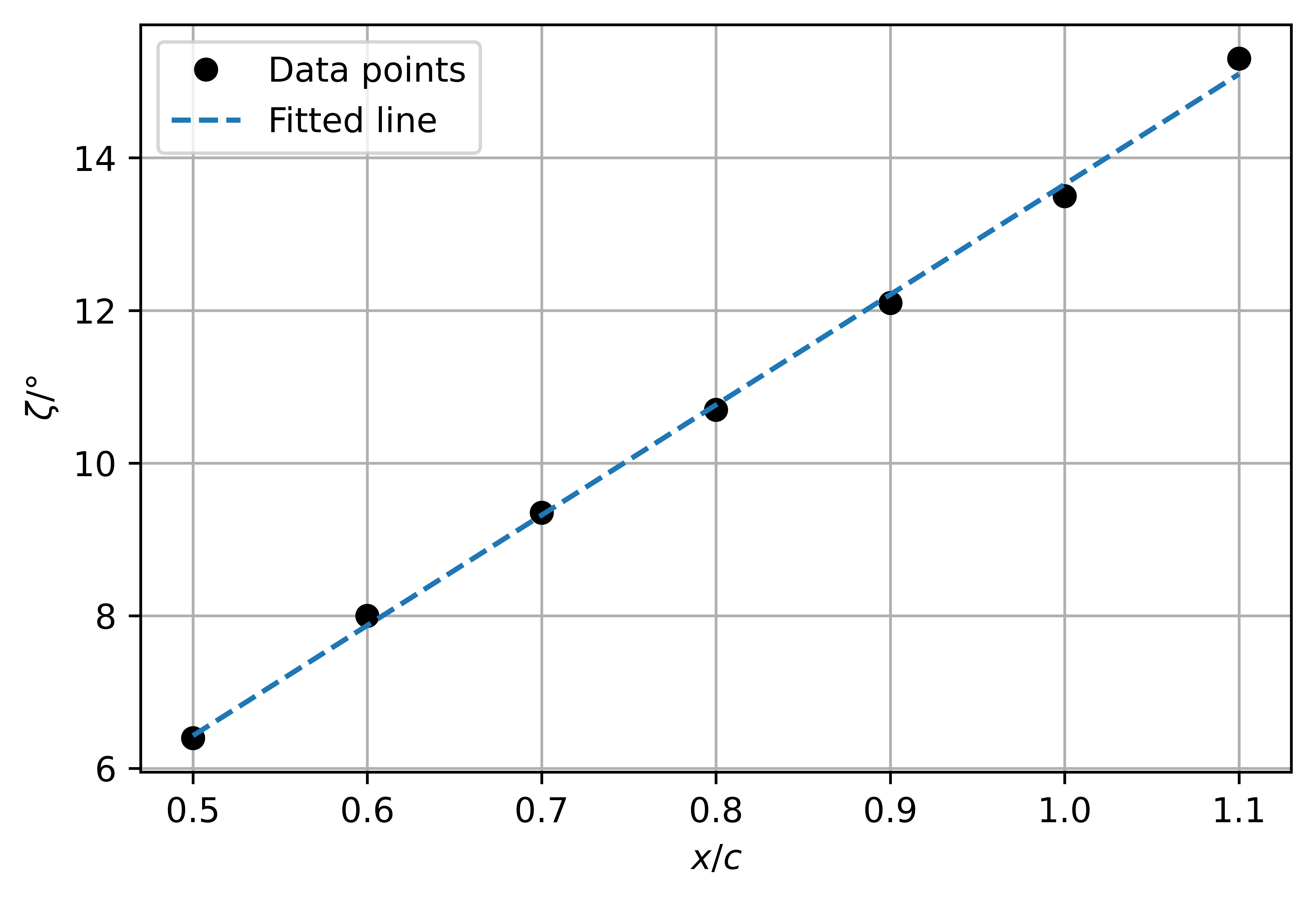}
\caption{Downstream evolution of the rotation angle $\zeta$ of the $w^*$ velocity stall cell structures.~The black points are the data points and the blue dashed line is the fitted line.}
\label{fig: Rotation of stall cells}
\end{figure}

For a given downstream distance, both positive and negative spanwise iso-contours rotate equally, with positive spanwise velocity iso-contours rotating clockwise and negative spanwise velocity iso-contours rotating counter-clockwise when viewed aligned with the inflow. Figure \ref{fig: Rotation of stall cells} shows the downstream evolution of this rotation angle, revealing a linear relationship with downstream distance. A linear fit yields the equation:

\begin{equation}
    \zeta = 14.5\left(\frac{x}{c}\right)-0.8.
    \label{eq:evolution equation of the stall cell rotation angle}
\end{equation}

This rotation of the spanwise velocity maxima can be attributed to the bending of the vortex sheet and vortex tubes. Around $z/c \approx \pm1$, the spanwise velocity is nearly zero for all downstream positions, and the comparatively low values of $w^*$ in these regions indicate reduced mixing, explaining the presence of the ``hollow" regions observed in the Q-criterion visualisations.

\begin{figure}
    \centering
    \includegraphics[width=\linewidth]{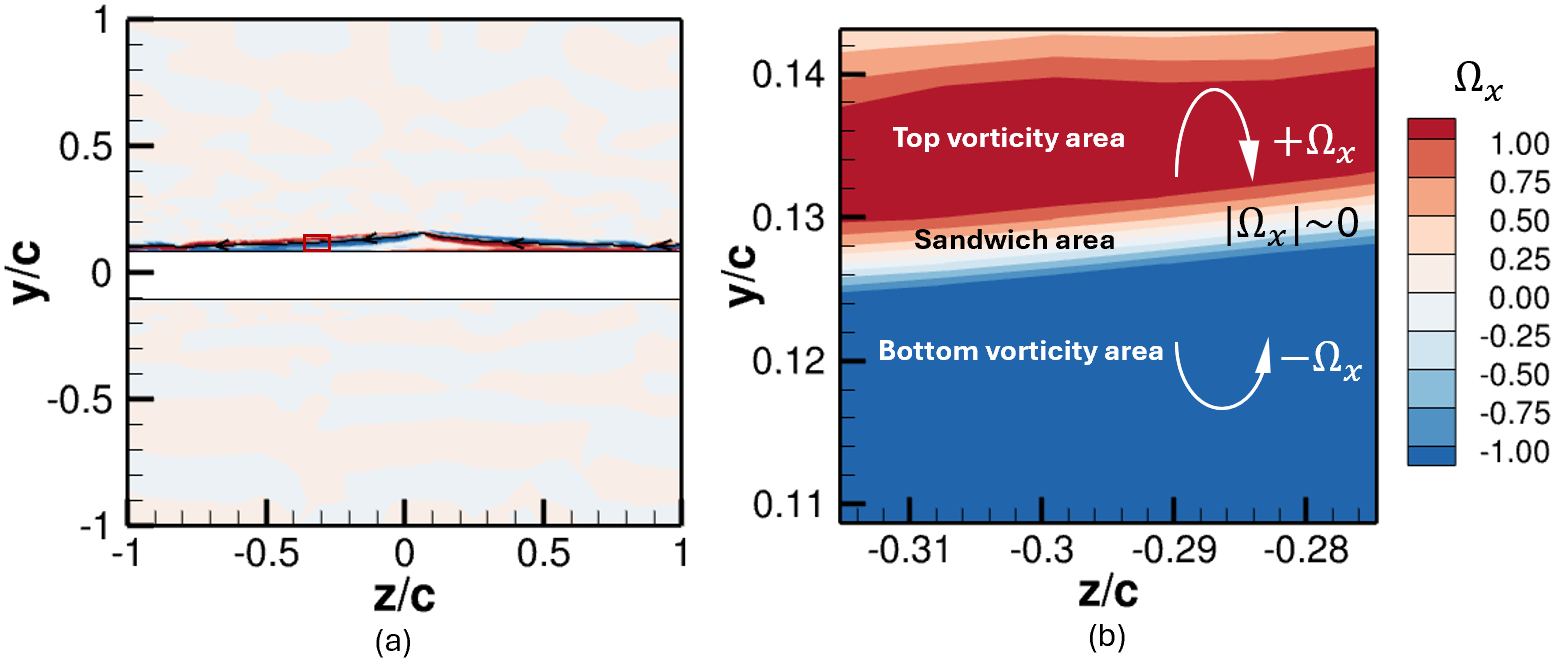}
    \caption{Contour plot of normalised $x$-vorticity $\Omega_x$ (a) at $x/c = 0.5$, illustrating different regions of $\Omega_x$. The enlarged view within the red box (b) shows the sandwich zone. Note that this is an enlarged view with $-1\leq y/c\leq 1$ and $-1\leq z/c\leq 1$ for the Figure (a).}
    \label{fig:different zones of x vorticity}
\end{figure}

The streamwise vorticity $\Omega_x$ also shows distinctive patterns in its downstream evolution. Figure \ref{fig:different zones of x vorticity} shows that $\Omega_x$ exhibits different regions: a Top Vorticity Area (TVA) and a Bottom Vorticity Area (BVA), separated by a Shear Area (SA) where $|\Omega_x|$ approaches zero. In the negative $z/c$ region, the BVA is negative and the TVA is positive, while in the positive $z/c$ region, the BVA is positive and the TVA is negative.

\begin{figure}
    \centering
    \includegraphics[width=\linewidth]{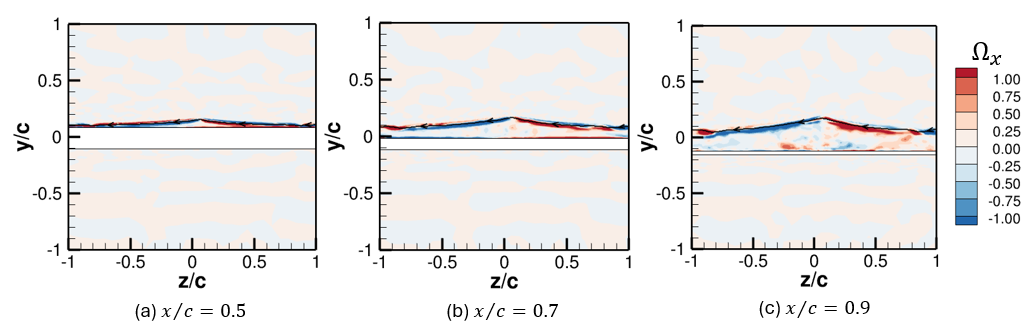}
    \caption{Contour plot of normalised  $x$-vorticity $\Omega_x$ at (a) $x/c$ = 0.5, (b) $x/c$ = 0.7, and (c) $x/c$ = 0.9 positions. Note that this is an enlarged view with $-1\leq y/c\leq 1$ and $-1\leq z/c\leq 1$.}
    \label{fig:logitudinal vorticity contour plot different down stream position}
\end{figure}

As the flow progresses downstream (Figure \ref{fig:logitudinal vorticity contour plot different down stream position}), the BVA increasingly dominates over the TVA. By $x/c = 0.9$, primarily the BVA remains. For visual clarity, zoomed versions of these contour plots within the region $-1 \leq y/c \leq 1$ and $-1 \leq z/c \leq 1$ are shown. This can be understood through the interaction of vortex sheets of opposite signs and unequal strengths. The BVA is visibly thicker than the TVA, and although the magnitude of $\Omega_x$ is similar in both regions at $x/c = 0.5$, the greater thickness of the BVA gives it greater vortex strength. As the flow progresses downstream and the vortex sheets interact, the stronger BVA eventually dominates and absorbs the weaker TVA, consistent with the vortex dynamics described by \cite{leweke2016dynamics}.\\


\subsection{Summary}

\noindent In summary, flow separation at the airfoil surface creates a low-pressure zone that drives recirculating flow. A shear layer (vortex sheet) develops at the interface between this recirculating region and the external downstream flow, forming a spanwise-oriented separation vortex tube characterised by $z$-vorticity, $\Omega_z$, which is the dominant vorticity component. Due to its proximity to the oppositely-rotating trailing edge vortex originating from the airfoil's pressure surface, the separation vortex tube undergoes Crow-type instability (see \cite{crow1970stability}). This instability mechanism deforms the vortex tube into a wave-like configuration. The resulting deformation propagates to the shear layer, producing a local protrusion near $z/c = 0$. These undulations in both the vortex tube and shear layer generate $y$-vorticity, $\Omega_y$, which maintains an approximately constant magnitude in the downstream direction.
The presence of the $\Omega_y$ component induces spanwise velocity $w$, whose vertical gradient in turn produces $x$-vorticity, $\Omega_x$. In the downstream region, BVA of $\Omega_x$ becomes dominant over TVA. The vertical velocity within the separation region is subsequently governed by the combined effects of $\Omega_x$ and $\Omega_z$. Fluid entrainment into the separated shear layer, driven by $\Omega_z$ vorticity, directs the entrained flow either in the $+z$ or $-z$ direction depending on the sign of $\Omega_y$, thereby reinforcing the instability through a self-sustaining mechanism. This three-dimensional flow structure ultimately manifests as a linear rotation pattern in the iso-contours of the normalised spanwise velocity component $w^*$.

\section{Effect of AoA } \label{sec:s8}

In the previous section, we examined the origin of stall cells through the vortex-based mechanism for results obtained at an AoA of 14°.  In this section, we examine the effect of increasing the angle of attack from 14° to 16°. We first analyse how this increase influences the number and spatial arrangement of stall cells. We then investigate the impact of the higher AoA on their downward evolution. Finally, we compare the temporal development of stall cells at 14° and 16° to assess whether the increase in AoA promotes stall cell instability.


\subsection{Effect on number of stall cells, and arrangement of stall cells}
\begin{figure}
\centering
\includegraphics[width=0.9\textwidth]{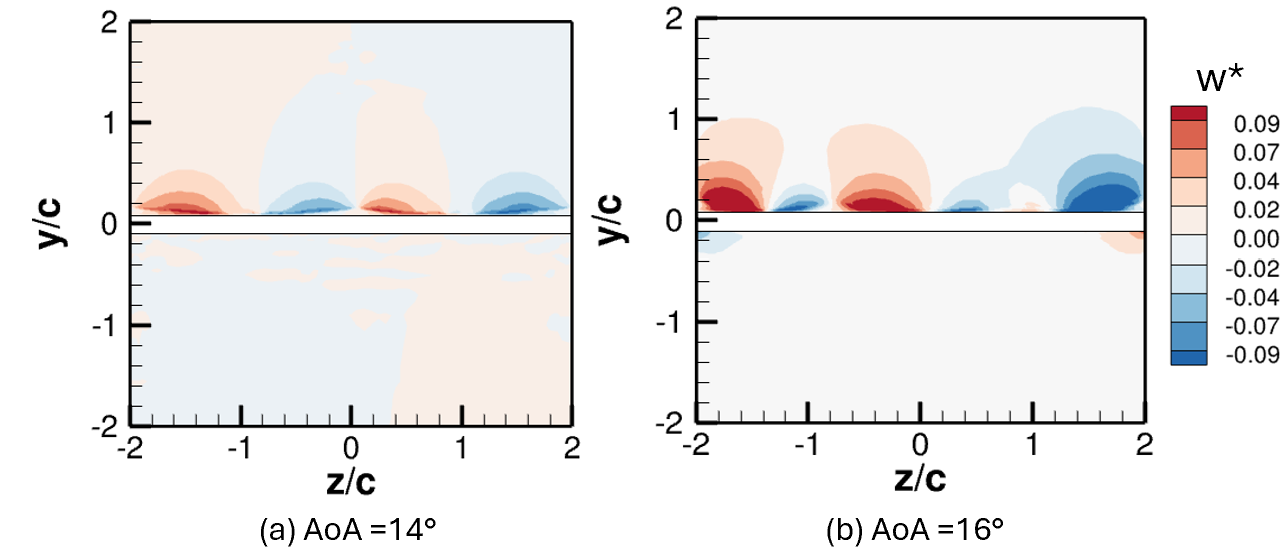}
\caption{Comparison of stall cell numbers observed, patterns, and arrangements at (a) 14° AoA and (b) 16° AoA using normalised mean $z$-velocity $w^*$ at $x/c = 0.5$.}
\label{fig:Comparison stall cells 16 deg vs 14 deg}
\end{figure}

Figure \ref{fig:Comparison stall cells 16 deg vs 14 deg} presents a comparison of the number, pattern, and arrangement of stall cells at $14^{\circ}$ and $16^{\circ}$ AoA, based on the normalised mean spanwise velocity $w^{*}$ at $x/c = 0.5$. With increase in AoA the number of stall cells increased from two at $14^{\circ}$ AoA to three at $16^{\circ}$ AoA. Furthermore, the spanwise velocity contours representing the stall cells exhibit a sign reversal about the line $z/c = 0$ as the AoA increases from $14^{\circ}$ to $16^{\circ}$ (see Figure \ref{fig:Comparison stall cells 16 deg vs 14 deg}). In addition, unlike the symmetric arrangement observed at $14^{\circ}$ AoA, the stall cells at $16^{\circ}$ AoA display an asymmetric distribution, with the region $z/c \in [-1,0]$ being more dominant than $z/c \in [0,1]$. This asymmetry in the mean flow suggests that, temporally, the stall cells at $16^{\circ}$ AoA are less stable compared to those at $14^{\circ}$ AoA. To validate this observation, Section \ref{sec:Temporal evolution of stall cells} presents a temporal analysis comparing the evolution of stall-cell arrangements between $14^{\circ}$ and $16^{\circ}$ AoA. 

\noindent Before going for the temporal analysis though, we will look at the downstream evolution of stall cells in the mean flow while comparing it with the evolution seen for 14° AoA (see section \ref{sec:Streamwise Evolution of the Vortex Core}).

\subsection{Effect on the downward evolution of stall cells}

\begin{figure}
    \centering
    \includegraphics[width=\linewidth]{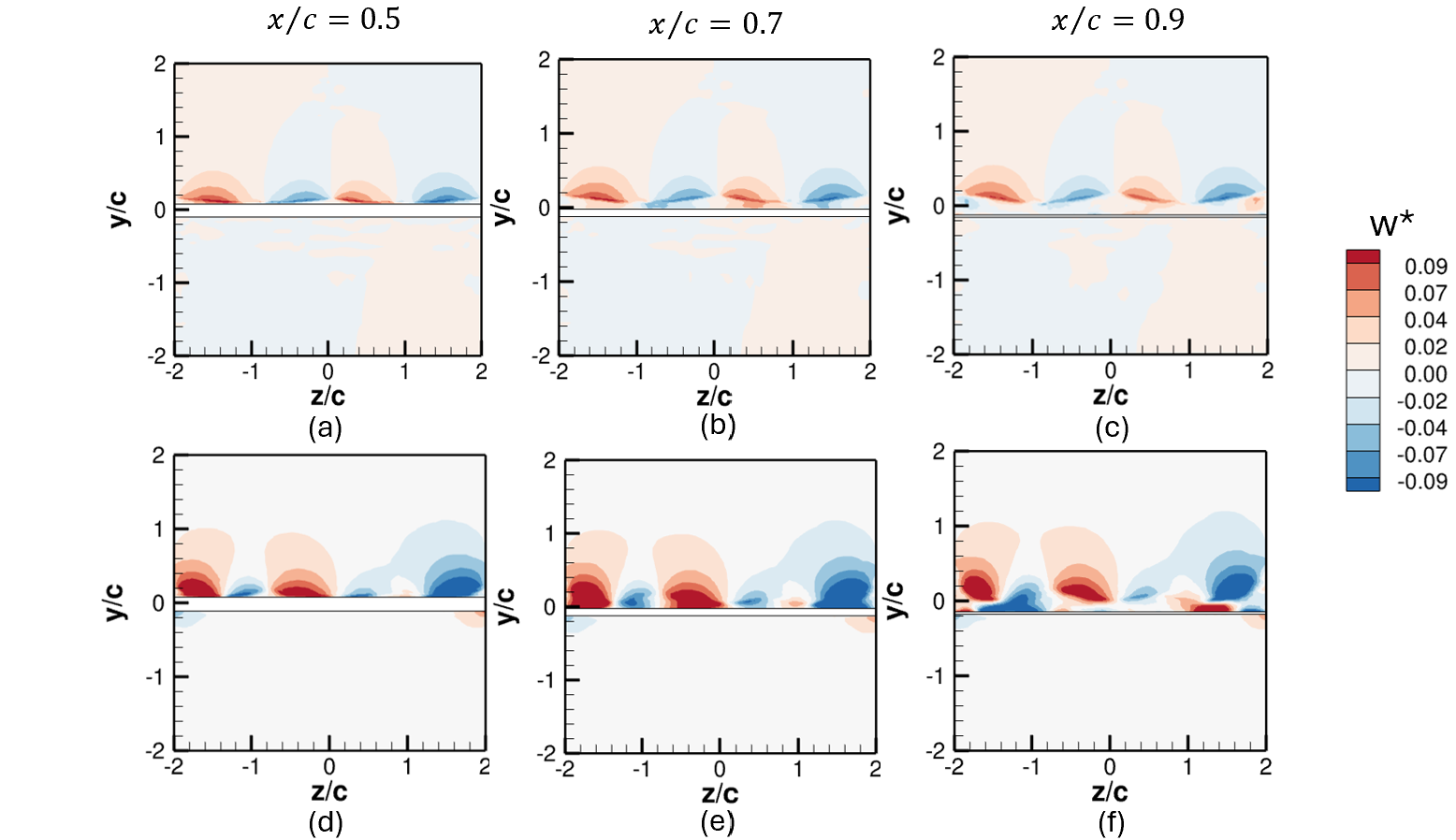}
    \caption{Comparison of contour plots of mean normalised $z$-velocity $w^*$ at $x/c = 0.5$, $x/c = 0.7$, and $x/c = 0.9$. The top row (a-c) corresponds to 14° AoA, and the bottom row (d-f) corresponds to 16° AoA.As the AoA shifts from $14^{\circ}$ to $16^{\circ}$, the spanwise velocity distribution within the stall cells undergoes a directional reversal about the line $z/c = 0$.}
    \label{fig:16 deg vs 14 deg downstream evolution comparison}
\end{figure}

Figure~\ref{fig:16 deg vs 14 deg downstream evolution comparison} compares the downstream evolution of stall cells at $14^{\circ}$ and $16^{\circ}$ AoA. The top row corresponds to $14^{\circ}$ AoA and the bottom row to $16^{\circ}$ AoA. As discussed in Section~\ref{sec:Streamwise Evolution of the Vortex Core}, for the $14^{\circ}$ AoA case, the spanwise velocity contours exhibit a rotation while largely preserving their shape as the flow progresses downstream. In contrast, for the $16^{\circ}$ AoA case, the downstream evolution of the stall cells is irregular, and the spanwise velocity contours progressively deform as the flow develops. This observation further indicates that the stall cells at $16^{\circ}$ AoA are temporally unstable compared with those at $14^{\circ}$ AoA.

\subsection{Temporal evolution of stall cells} \label{sec:Temporal evolution of stall cells}

\begin{figure}
    \centering
    \includegraphics[width=\linewidth]{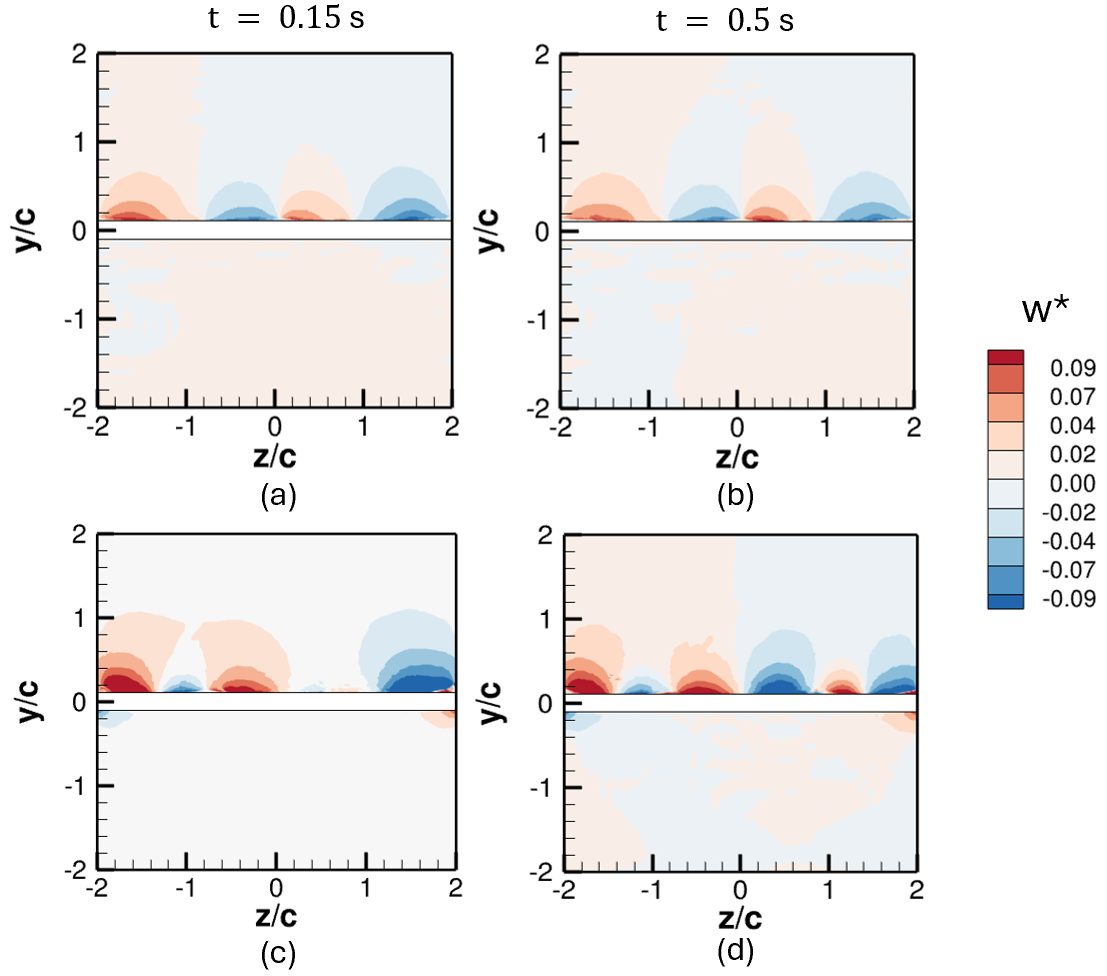}
    \caption{Comparison of contour plots of the normalised $z$-velocity $w^*$ at $x/c = 0.4$ for times (a) 0.15~s and (b) 0.7~s. The top row (a, b) corresponds to $14^\circ$ AoA, and the bottom row to $16^\circ$ AoA (c, d). At $16^\circ$ AoA, the stall cells, away from the wind tunnel wall, are predominantly located within $z/c \in [-1, 0]$ and are relatively small within $z/c \in [0, 1]$ at 0.15~s, whereas at 0.7~s this distribution becomes more symmetric around $z/c = 0$. In contrast, for $14^\circ$ AoA, away from the wind tunnel wall, stall cells remain distributed across both $z/c \in [-1, 0]$ and $z/c \in [0, 1]$ at both time instances, indicating stable stall-cell behaviour at $14^\circ$ AoA and unsteady or unstable behaviour at $16^\circ$ AoA.}
    \label{fig:16 deg vs 14 deg temporal evolution comparison temp}
\end{figure}

Figure~\ref{fig:16 deg vs 14 deg temporal evolution comparison temp} compares the contour plots of the normalised spanwise velocity $w^*$ at $x/c = 0.4$ for times (a) 0.15~s and (b) 0.7~s. The top row corresponds to $14^{\circ}$ AoA and the bottom row to $16^{\circ}$ AoA. At $16^{\circ}$ AoA, away from the wall, the stall cells are primarily concentrated within $z/c \in [-1, 0]$ and are comparatively weaker in $z/c \in [0, 1]$ at 0.15~s, whereas this distribution becomes more symmetric about $z/c = 0$ at 0.5~s. In contrast, for $14^{\circ}$ AoA, stall cells persist across both $z/c \in [-1, 0]$ and $z/c \in [0, 1]$ at both instants, indicating a temporally stable stall-cell structure. The observed alternation of stall-cell dominance at $16^{\circ}$ AoA suggests a loss of stability and a more unsteady stall-cell dynamics compared with the relatively steady behaviour at $14^{\circ}$ AoA.

\section{Conclusions} \label{sec:s9}

This study has provided a detailed investigation of the three-dimensional organisation and dynamics of stall cells over an airfoil at moderate Reynolds number. The key findings are summarised as follows:

\begin{enumerate}
\item \textbf{Global and local load characteristics}:
   \begin{itemize}
   \item The DDES-SST simulations accurately reproduce global aerodynamic loads, showing good agreement with validated RANS simulations.
   \item Pressure distributions reveal significant spanwise non-uniformity with symmetry about the mid-span, indicating that measurements at a single spanwise position are insufficient to characterise separated flows.
   \end{itemize}

\item \textbf{Wall friction and separation patterns}:
   \begin{itemize}
   \item Flow separation occurs non-uniformly along the span, with earliest separation at mid-span ($x/c = 0.28$) and delayed separation at $z/c \approx \pm1$ ($x/c = 0.4$) for 14° AoA.
   \item At mid-span, flow bifurcation reduces near-wall energy, promoting earlier separation, while flow convergence at $z/c \approx \pm1$ increases energy, delaying separation for 14° AoA.
   \end{itemize}

\item \textbf{Stall cell structure and identification}:
   \begin{itemize}
   \item Stall cells manifest as pairs of alternating regions of positive and negative spanwise velocity, with a wavelength of approximately 2c between $z/c = -1$ and $z/c = 1$ for 14° AoA.
   \item The Q-criterion visualisation reveals a wave-like pattern on the separated shear layer, with ``humps" at mid-span and near-wall regions and ``hollows" in between (see Figure \ref{fig:Q criterion isosurface 3D} (b)) for 14° AoA.
   \end{itemize}

\item \textbf{Vorticity dynamics and mechanism}:
   \begin{itemize}
   \item Flow separation generates a shear layer that culminates in a separation vortex tube with dominant spanwise vorticity ($\Omega_z$).
   \item This separation vortex tube interacts with the counter-rotating trailing edge vortex tube, experiencing a Crow-type instability that causes wave-like bending.
   \item The bending of the vortex tubes induces vertical vorticity ($\Omega_y$), which drives spanwise flow and contributes to the formation of stall cells.
   \item The vertical gradient of spanwise velocity dominates the generation of streamwise vorticity ($\Omega_x$).
   \end{itemize}

\item \textbf{Streamwise evolution}:
   \begin{itemize}
   \item The maxima of spanwise velocity structures rotate around fixed spanwise axes at $z/c = \pm1$ as the flow progresses downstream for 14° AoA.
   \item This rotation angle evolves linearly with downstream distance according to $\zeta = 14.5(x/c) - 0.8$ (for a range of $ 0.5\leq x/c \leq 1.1 $) a previously unreported quantitative relationship for 14° AoA.
   \item The streamwise vorticity field shows distinct regions namely top vorticity area (TVA) and bottom vorticity area (BVA) that evolve downstream, with the BVA eventually dominating due to its greater vortex strength for 14° AoA.
   \end{itemize}

\item \textbf{Increase in AoA}:
   \begin{itemize}
    \item As the AoA increases from 14° to 16°, the number of stall cells increases from 2 to 3. Unlike the 14° case, an asymmetry about $z/c = 0$ is observed in the stall cell arrangement for the 16° case.
    \item Unlike in the 14° case, where spanwise velocity contours rotate while preserving their shape in the downstream direction, the downstream evolution in the 16° case is irregular, and the contours deform progressively.
    \item For the 14° AoA case, the spanwise velocity contours are stable in time, while for the 16° AoA case, an alteration in stall cell dominance is observed, suggesting a loss of stability and more unstable stall cell dynamics.
\end{itemize}

\end{enumerate}

\noindent These findings enhance our understanding of the three-dimensional nature of separated flows over airfoils and provide new insights into the vorticity mechanisms underlying stall cell formation and evolution. The linear relationship governing the rotation of spanwise velocity structures is particularly significant as it offers a quantitative description of stall cell evolution that could inform future predictive models.

\noindent This study has shown that stall cells form even under moderately high inflow turbulence conditions (3\%), suggesting their relevance to practical applications like wind turbine blades operating in turbulent environments. The non-uniform spanwise load distribution associated with stall cells has important implications for structural loading and aerodynamic performance predictions.

\noindent Future work should investigate the generality of these findings across different angles of attack, Reynolds numbers, aspect ratios and airfoil geometries. Additional studies exploring the interaction between stall cells and dynamic effects, such as blade rotation or pitch oscillations, would further enhance our understanding of these complex three-dimensional flow structures and their impact on aerodynamic performance.\\

The vorticity dynamics elucidated in this work, the Crow-type instability 
of the counter-rotating separation and trailing-edge vortex tubes, the 
wave-like bending of the vortex sheet, and the generation of $\Omega_y$ 
driving alternating spanwise velocity, provide a clear physical picture 
of stall cell formation. These observations motivate and guide the 
development of an analytical model presented in a companion paper \citep{PartII}. In that work, the linear stability analysis of the coupled vortex tube system yields the instability growth rate and wavelength selection, while a weakly nonlinear analysis using the method of multiple scales derives the Stuart--Landau amplitude equation governing saturation. The vortex sheet is coupled to the filament dynamics through the Birkhoff--Rott equation, from which the vertical vorticity and spanwise velocity fields are derived analytically. The model predictions for vortex sheet deformation, $\Omega_y$ distribution, and spanwise velocity magnitude are validated against the DDES results presented in this paper, providing a unified theoretical framework for stall cell formation.

\begin{bmhead}[Author ORCIDs]{R. Mishra, https://orcid.org/0000-0001-9727-7281 ; E. Guilmineau, https://orcid.org/0000-0001-9070-093X; I. Neunaber, https://orcid.org/0000-0002-3787-3118; C. Braud, https://orcid.org/0000-0002-4409-4278.}
\end{bmhead}

\begin{bmhead}[Author contributions]{
\textbf{R.M.}:
Investigation; Methodology; Formal analysis; Validation; Visualisation; Writing: original draft. \textbf{E.G.}: Methodology; Writing: review \& editing, Supervision; Acquisition of the HPC grant. \textbf{I.N.}: Writing: review \& editing, Supervision. \textbf{C.B.}: Conceptualisation, Funding Acquisition, Supervision.}
\end{bmhead}

\begin{bmhead}[Funding]
This work is funded under French national project MOMENTA (grant no. ANR-19-CE05-0034). The author also gratefully acknowledges the financial support provided by CARNOT MER's LIFEMONITOR project. All simulation in this work is supported by computing HPC and storage resources by
GENCI at IDRIS thanks to the grant 2024-A0172A13014and
2025-A0192A13014on the supercomputer Jean Zay's CSL partition.
\end{bmhead}

\begin{bmhead}[Declaration of interests]
The authors report no conflict of interest.
\end{bmhead}

\bibliographystyle{jfm}
\bibliography{stallcells}

\end{document}